\font\bbb=msbm10                                                   %%%
%\def\bbb{\bf}                                                     %%%
%%%                                                                %%%
%%%                                                                %%%
%%%%%%%%%%%%%%%%%%%%%%%%%%%%%%%%%%%%%%%%%%%%%%%%%%%%%%%%%%%%%%%%%%%%%%
%%%%%%%%%%%%%%%%%%%%%%%%%%%%%%%%%%%%%%%%%%%%%%%%%%%%%%%%%%%%%%%%%%%%%%

\def\C{\hbox{\bbb C}}

\def\Z{\hbox{\bbb Z}}

\def\ANYAS{{\sl Ann.\ NY Acad.\ Sci.}}

\def\BSTJ{{\sl Bell System Tech.\ J.}}

\def\CMP{{\sl Commun.\ Math.\ Phys.}}

\def\CS{{\sl Complex Systems}}

\def\FP{{\sl Found.\ Phys.}}

\def\IJTP{{\sl Int.\ J. Theor.\ Phys.}}
\def\IPL{{\sl Info.\ Processing Lett.}}

\def\JAP{{\sl J. Appl.\ Phys.}}
\def\JCSS{{\sl J. Comp.\ Syst.\ Sci.}}
\def\JLMS{{\sl Proc.\ Lond.\ Math.\ Soc.}}

\def\JSP{{\sl J. Stat.\ Phys.}}
\def\MST{{\sl Math.\ Systems Theory}}

\def\PAMS{{\sl Proc.\ Amer.\ Math.\ Soc.}}

\def\PD{{\sl Physica D}}

\def\PNAW{{\sl Proc.\ Nederl.\ Akad.\ Wetensch.}}

\def\PRA{{\sl Phys.\ Rev.\ A}}

\def\PRL{{\sl Phys.\ Rev.\ Lett.}}
\def\PRSLA{{\sl Proc.\ Roy.\ Soc.\ Lond.\ A}}
\def\PSAM{{\sl Proc.\ Symp.\ Appl.\ Math.}}

\def\TIEICEJE{{\sl Trans.\ IEICE Japan E}}

\def\dajm{\hbox{D. A. Meyer}}

\def\rds{\hbox{R. D. Sorkin}}

\def\brosl{\hbox{B. Hasslacher}}

\def\feynman{\hbox{R. P. Feynman}}
\def\deutsch{\hbox{D. Deutsch}}
\def\teich{\hbox{W. G. Teich}}
\def\gz{\hbox{G. Gr\"ossing and A. Zeilinger}}

\def\hfb{\hfil\break}

\catcode`@=11
\newskip\ttglue

   \font\ninerm=cmr9    \font\eightrm=cmr8   \font\sixrm=cmr6
  \font\ninebf=cmbx9   \font\eightbf=cmbx8  \font\sixbf=cmbx6
  \font\nineit=cmti9   \font\eightit=cmti8  
  \font\ninesl=cmsl9   \font\eightsl=cmsl8  
  \font\ninemi=cmmi9   \font\eightmi=cmmi8  \font\sixmi=cmmi6

\font\bigtenbf=cmr10 scaled\magstep2 

\def\ninepoint{\def\rm{\fam0\ninerm}%
  \textfont0=\ninerm \scriptfont0=\sixrm
  \textfont1=\ninemi \scriptfont1=\sixmi
  \textfont\itfam=\nineit  \def\it{\fam\itfam\nineit}%
  \textfont\slfam=\ninesl  \def\sl{\fam\slfam\ninesl}%
  \textfont\bffam=\ninebf  \scriptfont\bffam=\sixbf
    \def\bf{\fam\bffam\ninebf}%
  \tt \ttglue=.5em plus.25em minus.15em
  \normalbaselineskip=11pt
  \setbox\strutbox=\hbox{\vrule height8pt depth3pt width0pt}%
  \normalbaselines\rm}

\def\eightpoint{\def\rm{\fam0\eightrm}%
  \textfont0=\eightrm \scriptfont0=\sixrm
  \textfont1=\eightmi \scriptfont1=\sixmi
  \textfont\itfam=\eightit  \def\it{\fam\itfam\eightit}%
  \textfont\slfam=\eightsl  \def\sl{\fam\slfam\eightsl}%
  \textfont\bffam=\eightbf  \scriptfont\bffam=\sixbf
    \def\bf{\fam\bffam\eightbf}%
  \tt \ttglue=.5em plus.25em minus.15em
  \normalbaselineskip=9pt
  \setbox\strutbox=\hbox{\vrule height7pt depth2pt width0pt}%
  \normalbaselines\rm}

\def\sfootnote#1{\edef\@sf{\spacefactor\the\spacefactor}#1\@sf
      \insert\footins\bgroup\eightpoint
      \interlinepenalty100 \let\par=\endgraf
        \leftskip=0pt \rightskip=0pt
        \splittopskip=10pt plus 1pt minus 1pt \floatingpenalty=20000
        \parskip=0pt\smallskip\item{#1}\bgroup\strut\aftergroup\@foot\let\next}
\skip\footins=12pt plus 2pt minus 2pt
\dimen\footins=30pc

\def\ie{{\it i.e.}}
\def\eg{{\it e.g.}}

\def\etal{{\it et al.}}

\def\Tr{{\hbox{Tr}}}
\def\Proof{{\sl Proof}}
\def\Lemma{L{\eightpoint EMMA}}
\def\Theorem{T{\eightpoint HEOREM}}
\def\Corollary{C{\eightpoint OROLLARY}}

\def\endproof{\vrule height6pt width6pt depth0pt}
\def\llangle{\langle\!\langle}
\def\rrangle{\rangle\!\rangle}

\def\Gone{G_1(Q,k)}
\def\Gtwo{G_2(Q,k)}
\def\Done{D_1(Q,k,f)}
\def\Dtwo{D_2(Q,k,f)}

\magnification=1200
\input epsf.tex

\dimen0=\hsize \divide\dimen0 by 13 \dimendef\chasm=0
\dimen1=\hsize \advance\dimen1 by -\chasm \dimendef\usewidth=1
\dimen2=\usewidth \divide\dimen2 by 2 \dimendef\halfwidth=2
\dimen3=\usewidth \divide\dimen3 by 3 \dimendef\thirdwidth=3
\dimen4=\hsize \advance\dimen4 by -\halfwidth \dimendef\secondstart=4
\dimen5=\halfwidth \advance\dimen5 by -10pt \dimendef\indenthalfwidth=5
\dimen6=\thirdwidth \multiply\dimen6 by 2 \dimendef\twothirdswidth=6
\dimen7=\twothirdswidth \divide\dimen7 by 4 \dimendef\qttw=7
\dimen8=\qttw \divide\dimen8 by 4 \dimendef\qqttw=8
\dimen9=\qqttw \divide\dimen9 by 4 \dimendef\qqqttw=9

\parskip=0pt
\line{\hfil February 1996}
\line{\hfil {\it revised\/} April 1996}
\line{\hfil quant-ph/9605023}
\bigskip\bigskip\bigskip
\centerline{\bf\bigtenbf UNITARITY IN ONE DIMENSIONAL}
\bigskip
\centerline{\bf\bigtenbf NONLINEAR QUANTUM CELLULAR AUTOMATA}
\vfill
\centerline{\bf David A. Meyer}
\bigskip 
\centerline{\sl Project in Geometry and Physics}
\centerline{\sl Department of Mathematics}
\centerline{\sl University of California/San Diego}
\centerline{\sl La Jolla, CA 92093-0112}
\centerline{dmeyer@euclid.ucsd.edu}
\vfill
\centerline{ABSTRACT}
\bigskip
%--------|---------|---------|---------|---------|---------|---------|
\noindent Unitarity of the global evolution is an extremely stringent 
condition on finite state models in discrete spacetime.  Quantum
cellular automata, in particular, are tightly constrained.  In 
previous work we proved a simple No-go Theorem which precludes 
nontrivial homogeneous evolution for linear quantum cellular automata.  
Here we carefully define general quantum cellular automata in order to 
investigate the possibility that there be nontrivial homogeneous 
unitary evolution when the local rule is nonlinear.  Since the unitary 
global transition amplitudes are constructed from the product of local 
transition amplitudes, infinite lattices require different treatment 
than periodic ones.  We prove Unitarity Theorems for both cases, 
expressing the equivalence in $1+1$ dimensions of global unitarity and 
certain sets of constraints on the local rule, and then show that 
these constraints can be solved to give a variety of multiparameter 
families of nonlinear quantum cellular automata.  The Unitarity 
Theorems, together with a Surjectivity Theorem for the infinite case, 
also imply that unitarity is decidable for one dimensional cellular 
automata.

\bigskip
%--------|---------|---------|---------|---------|---------|---------|

\vfill
\eject

\headline{\ninepoint\it Unitarity in nonlinear QCAs
          \hfil David A. Meyer}
\parskip=10pt

\noindent{\bf 1.  Introduction}

%--------|---------|---------|---------|---------|---------|---------|
\noindent Already present in Feynman's inspirational essay on quantum
physics and computation [1] is the concept of a binary quantum 
cellular automaton (QCA):  a discrete spacetime array of quantum
processors, each of which has two eigenstates (un/occupied or spin 
up/down) and is coupled to some set of neighboring processors.  He 
explicitly recognizes the difficulties of reconciling discreteness and
locality of interaction with Lorentz invariance---the same problem
which must be solved in the causal set approach to quantum gravity 
[2].  Gr\"ossing, Zeilinger, \etal, discovered a similar conflict with
group invariance in their attempts to apply QCAs as quantum 
simulators, \ie, for quantum rather than deterministic computation.%
\sfootnote*{Feynman [3], Margulis [4], and more recently, Lent and 
Tougaw [5] have investigated the possiblilities for deterministic 
computation using QCAs.}
Being unable to reconcile discreteness and locality with translation 
invariance and unitarity, they were led to study a class of CAs whose 
evolution is, although `probability' preserving, both nonunitary [6] 
and nonlocal [7].

%--------|---------|---------|---------|---------|---------|---------|
Near the beginning of our investigation into exactly unitary---and 
therefore truly quantum---CAs we proved that these physically 
reasonable requirements are actually incompatible for this class of 
CAs.  More precisely, taking the definition of QCA to include 
unitarity, discreteness and locality, we proved [8]:

%--------|---------|---------|---------|---------|---------|---------|
\noindent N{\eightpoint O-GO} \Theorem.  {\sl No nontrivial 
homogeneous linear binary QCA exists on Euclidean lattices in any
dimension.}

%--------|---------|---------|---------|---------|---------|---------|
We showed, however, that weakening the homogeneity condition slightly
to require invariance under only a subgroup of translations allows the
existence of linear binary QCAs [9].  The simplest of these 
{\sl partitioned\/} [10] QCAs models the evolution of a quantum 
particle and, in the appropriate limit, simulates the $1+1$ 
dimensional Dirac equation.  This physical interpretation motivates 
two reformulations of the QCA [9]:

%--------|---------|---------|---------|---------|---------|---------|
\item{}As a quantum lattice gas automaton.  The QCA may be formulated
       as a lattice gas with a single particle.  Generalizing to 
       multiple particles forces the automaton to be nonlinear if we 
       impose an exclusion principle.  The dimension of the Hilbert 
       (Fock) space of the theory is now exponential in the 
       cardinality of the lattice.  If there is no particle 
       creation/annihilation, the one particle sector of the Fock 
       space is equivalent to the original QCA.

%--------|---------|---------|---------|---------|---------|---------|
\item{}As a homogeneous linear ternary QCA.  In the Dirac equation the 
       amplitudes for the particle to be left/right moving at a point 
       are combined into a two component field.  Equivalently, the QCA 
       may be formulated to have three eigenstates at each cell, 
       corresponding to empty, left moving, and right moving.
       Coupling two copies of the original QCA makes the new one 
       homogeneous.

\noindent Both generalizations evade the No-go Theorem; the first by
being both nonlinear and not quite homogeneous and the second by being
ternary.

%--------|---------|---------|---------|---------|---------|---------|
Emulating Morita and Harao's construction of a computation universal
reversible one dimensional CA [11], Watrous has recently constructed a
one dimensional QCA which is quantum computation universal [12] in the
sense that it efficiently simulates the universal quantum Turing 
machine defined by Bernstein and Vazirani [13], which in turn 
efficiently simulates any quantum Turing machine as originally defined 
by Deutsch [14].  Both Morita and Harao's and Watrous' universal CAs 
are homogeneous, but since each may be considered to consist of three 
coupled partitioned CAs,%
\sfootnote*{Morita subsequently constructed a simpler universal 
reversible one dimensional CA which comprises two coupled partitioned
CAs [15]; presumably the analogous construction also works in the
quantum context.}
they may also be described as partitioned.  They are not, therefore,
binary---far more eigenstates are required.  Since a primary 
motivation for considering QCA models for quantum computation is the 
likelihood that massive parallelism will optimize nanoscale computer 
architecture [16] and the most plausible nanoscale devices have only a 
few eigenstates [17], it is of interest to ask if the No-go Theorem 
may also be evaded by homogeneous {\sl nonlinear\/} binary QCAs.  In 
this paper we answer that question affirmatively and, as a first step
toward exploring the computational power of such architectures, we 
explicitly parameterize the rule spaces for the simplest such QCAs.

%--------|---------|---------|---------|---------|---------|---------|
We begin in Section 2 by carefully defining what it means for a CA to
be quantum mechanical.  Rather than restricting a QCA to have only 
finite configurations as do Watrous [12] and D\"urr, Thanh and Santha 
[18], we consider both periodic and infinite lattices.  In order for 
the global transition amplitudes to be well defined as the product of 
local transition amplitudes we must distinguish more carefully between 
these two situations than is necessary in the deterministic case.  The 
distinction is emphasized in the following section by the observation 
that infinite QCAs must be asymptotically deterministic, though not 
necessarily quiescent.

%--------|---------|---------|---------|---------|---------|---------|
In Section 3 we prove a series of results expressing the equivalence
of unitary global evolution and sets of constraints on the local 
transition amplitudes.  Utilizing a bijection between 
configurations/pairs of configurations and paths on weighted graphs
$G_1$/$G_2$ we prove Unitarity Theorems 3.9 and 3.10.  These show that
unitarity is equivalent to sets of constraints on the weights of such
paths arising from the condition that the global evolution be norm
preserving and, in the infinite case, from the additional independent 
condition that the global evolution be surjective; the latter is the 
content of Surjectivity Theorem 3.18.

%--------|---------|---------|---------|---------|---------|---------|
We observe in Section 4 that the Unitarity and Surjectivity Theorems, 
together with the finiteness of the graphs $G_1$ and $G_2$, prove that 
unitarity is decidable in one dimension.  D\"urr and Santha have
obtained the same result by different methods [19].  Our main 
interest in this section, however, is to extract from these theorems a 
procedure for finding multiparameter families of local rules which 
define QCAs.  We do so, and apply it in the cases where the local 
neighborhood has size 2 or 3.

%--------|---------|---------|---------|---------|---------|---------|
At the beginning of Section 5 we show that a pattern observed in the 
derivation of the small local neighborhood rules in Section 4 
generalizes to give families of QCAs for any size local neighborhoods.
We conclude by discussing connections with deterministic reversible
CAs and point in directions for further research to explore the
computational power and physical interpretation of nonlinear QCAs.

\medskip
\noindent{\bf 2.  Definitions}
\nobreak

\nobreak
%--------|---------|---------|---------|---------|---------|---------|
\noindent A homogeneous CA is defined by a 4-tuple $(L,Q,f,E)$:  For 
the purposes of this paper we will take the {\sl lattice of cells\/} 
$L$ to be the integers $\Z$ (possibly with periodic identifications to 
$\Z_N$); such a CA is {\sl one dimensional}.  $Q$ is a finite set of 
{\sl states\/} $\{0,\ldots,q-1\}$; {\sl configurations\/} are maps 
$\sigma : L \to Q$, the argument of which will be denoted by a 
subscript.  The {\sl local rule\/} $f : Q \times Q^k \to \C$ defines 
the dynamics of the CA which it will be convenient to encode as a set 
of {\sl amplitude vectors\/}:  for each {\sl local configuration\/} 
$\lambda = (i_1,\ldots,i_k) \in Q^k$,
$$
|\lambda\rrangle :=
\bigl(f(0|i_1,\ldots,i_k),\ldots,f(q-1|i_1,\ldots,i_k)\bigr) \in \C^q.
$$
We use the variation $|\cdot\rrangle$ of the familiar Dirac notation 
[20] to denote a vector in $\C^q$ while distinguishing it from a state 
vector of some quantum system.  The sesquilinear inner product on 
$\C^q$ is denoted $\llangle\cdot|\cdot\rrangle$.  Also, the notation 
for the arguments of $f$ has been chosen to evoke that of conditional 
probability so that $f(i|\lambda)$ is the $i^{\hbox{\eightpoint th}}$ 
component in the amplitude vector, given that $\lambda$ is the local 
configuration.

%--------|---------|---------|---------|---------|---------|---------|
If the range of $f$ is $\{0,1\} \subset \C$ and each amplitude vector 
has exactly one nonzero component, the CA is {\sl deterministic\/}:  
there is a {\sl global evolution map\/} $F : L^Q \to L^Q$ defined by
$$
\sigma' = F \sigma 
\quad \iff \quad 
\forall x \in L,\; f(\sigma'_x | \sigma^{\vphantom\prime}_{x+E}) = 1,
                                                            \eqno(2.1)
$$
where the {\sl local neighborhood\/} $E := \{e_1,\ldots,e_k\}$ is a 
finite set of lattice vectors which defines the 
{\sl $E$-subconfigurations\/} 
$\sigma_{x+E} := (\sigma_{x+e_1},\ldots,\sigma_{x+e_k}) \in Q^k$.  $F$ 
is well defined since for each $E$-subconfiguration $\sigma_{x+E}$ 
there is a unique $\sigma'_x$ such that 
$f(\sigma'_x|\sigma^{\vphantom\prime}_{x+E}) = 1$.  It is convenient 
to take the local neighborhood to be connected; this is no loss of 
generality since the amplitude vectors can be independent of part of 
the local configuration.  In one dimension this means the local 
neighborhood is a sequence of $k$ consecutive integers and local 
configurations can be written as strings $i_1\ldots i_k$, $i_j \in Q$, 
indicating states of consecutive cells.

%--------|---------|---------|---------|---------|---------|---------|
When any of the amplitude vectors has more than a single nonzero 
component the CA is called {\sl indeterministic\/} [21]; probabilistic
CA models for parallel computation, considered already in the original
work of von Neumann and Ulam [22], fall into this class.  Here we are
interested in the quantum mechanical situation so the values of $f$ 
are probability amplitudes and the state of the CA is described at 
each timestep by a {\sl configuration vector\/} 
$\phi(t) \in \C L^Q$, the complex vector space with a basis 
$\{|\sigma\rangle \mid \sigma \in L^Q\}$ labelled by the 
configurations.  The global evolution map $F : \C L^Q \to \C L^Q$ is 
defined on the configuration basis by
$$
F |\sigma\rangle := 
 \sum_{\sigma'\in L^Q} F_{\sigma'\sigma} |\sigma'\rangle,  \eqno(2.2a)
$$
where
$$
F_{\sigma'\sigma} := 
 \prod_{x\in L} 
  f(\sigma'_x|\sigma^{\vphantom\prime}_{x+E}),             \eqno(2.2b)
$$
and then extended linearly to all of the {\sl configuration space\/}
$\C L^Q$.  Although the global evolution is linear on the 
configuration space, the local dynamics of the CA is {\sl nonlinear\/} 
unless the local rule is constant or additive [23], \ie, unless the 
amplitude vector is a linear function of the local configuration.

%--------|---------|---------|---------|---------|---------|---------|
When the CA is deterministic, each factor in the product $(2.2b)$ is 
either 0 or 1.  If any is 0 the product is 0, otherwise it is 1.  Thus 
definition (2.2) subsumes definition (2.1).

%--------|---------|---------|---------|---------|---------|---------|
When the CA is indeterministic, however, we must be more precise about
the meaning of the product in $(2.2b)$.  There are two possiblities:

%--------|---------|---------|---------|---------|---------|---------|
If $L = \Z_N$ there is no problem.  All the {\sl transition 
amplitudes\/} $F_{\sigma'\sigma}$ are defined by finite products.  To 
study the CA on an infinite lattice, we may take the usual statistical 
mechanics approach of computing some property (\eg, a correlation 
function) on a sequence of lattices of increasing size and 
investigating the limiting behaviour of that property.  With this 
methodology in mind we refer to a family of CAs $(\Z_N,Q,f,E)$ for all 
positive integers $N$ as a {\sl periodic\/} CA and call a 
configuration {\sl admissible\/} for a periodic CA if it is periodic 
with any finite period $N$.

%--------|---------|---------|---------|---------|---------|---------|
If $L = \Z$, the infinite product $(2.2b)$ is defined as the limit of 
a sequence of partial products.  A nonzero limit only exists if the 
successive factors in the product converge to 1.  Since the range of 
$f$ is at most $q^{k+1}$ points in $\C$, this can only happen if 
beyond some stage all the factors in the product are 1.   For any
configuration $\sigma$ define the {\sl limit set}
$$
\Omega^+(\sigma) := 
 \{\lambda \in Q^k \mid 
   \forall R > 0, \exists x > R : \sigma_{x+E} = \lambda
 \},
$$
and define $\Omega^-(\sigma)$ similarly, \ie, by conditioning on the 
existence of $x < -R$.  We will refer to the two maximal 
semi-infinite subconfigurations of $\sigma$ which contain only 
$E$-subconfigurations from the respective limit set as the 
{\sl ends\/} of $\sigma$.  Then $\sigma$ has a nonzero transition 
amplitude to some configuration only if for each 
$\lambda \in \Omega^{\pm}(\sigma)$ some component of the amplitude 
vector $|\lambda\rrangle$ is 1.  For any local rule $f$, let
$$
B_{\! f} := 
 \{\lambda \in Q^k \mid 
   \exists i \in Q : f(i|\lambda) = 1
 \},
$$
be the set of local configurations with {\sl big\/} amplitude vectors.
This description is justified by the observation that for each 
$\lambda \in B_{\! f}$, the length of its amplitude vector is greater
than or equal to 1.  Call a set of configurations $A \subset \Z^Q$ 
{\sl admissible\/} iff
\parskip=0pt
\medskip
\itemitem{  ({\it i\/})}$\forall\sigma,\sigma'\in A$, 
                        $F_{\sigma'\sigma}$ is well defined.
\itemitem{ ({\it ii\/})}$\forall\sigma\in A$, $\exists\sigma'\in A$
                        such that $F_{\sigma'\sigma}\not=0$.
\itemitem{({\it iii\/})}$\forall\sigma\in A$, $\sigma'\in\Z^Q$, if
                        $F_{\sigma'\sigma}\not=0$ then $\sigma'\in A$.
\medskip
\noindent We allow {\sl infinite\/} CAs only when the set of 
admissible configurations is nonempty.  Several simple facts follow
immediately:
\parskip=10pt

%--------|---------|---------|---------|---------|---------|---------|
\noindent\Lemma\ 2.1.  {\sl $A$ is closed under the evolution.
Furthermore, if $\sigma \in A$ then\/
$\Omega^{\pm}(\sigma) \subset B_{\! f}$, and if\/ 
$\Omega^{\pm}(\sigma') = \Omega^{\pm}(\sigma)$ then $\sigma' \in A$.}

%--------|---------|---------|---------|---------|---------|---------|
The most familiar infinite CAs have a unique {\sl quiescent\/} state 
[22], say 0, defined by the property that 
$f(i|0\ldots 0) = \delta_{i0}$, \ie, $B_{\! f} = \{0\ldots 0\}$.  
(These are the only QCAs allowed in [12] and [18,19].)  For such CAs 
admissible configurations have {\sl finite support\/} in the sense 
that the only nonquiescent cells lie in some finite domain.  It is 
clear that the collection of configurations with finite support is 
closed under the global evolution and, depending on $f$, may satisfy 
the admissibility conditions ({\it i\/})--({\it iii\/}); in Section 3 
we show that these properties can hold for more interesting sets of
configurations as well.

%--------|---------|---------|---------|---------|---------|---------|
Consider the subspace of $\C L^Q$ with basis the admissible 
configurations $A$.  This will become a Hilbert space, the 
{\sl physical configuration space\/} $H$, once we define an inner 
product.  Again using Dirac notation [20], but now for vectors 
$|\phi\rangle,|\psi\rangle \in \C A$ denoting states of a quantum 
system,
$$
\langle \phi|\psi \rangle := 
 \sum_{\sigma\in A} \overline\phi_{\sigma} 
             \psi^{\vphantom 1}_{\sigma},
$$
where 
$|\phi\rangle = \sum_{\sigma\in A} \phi_{\sigma} |\sigma\rangle$ and 
similarly for $|\psi\rangle$.  $H \subset \C A$ is the set of vectors
with finite norm.  A periodic or infinite CA is a {\sl quantum\/} CA 
if the global evolution preserves this inner product and is time 
reversible (which is a distinct condition for infinite dimensional 
Hilbert space [24]), \ie, is {\sl unitary}.  The physical 
interpretation is that when the system is described by the 
configuration vector $|\phi\rangle$ with 
$\langle\phi|\phi\rangle = 1$, the probability of observing it to be 
in configuration $\sigma$ is 
$\overline\phi_{\sigma}\phi^{\vphantom 1}_{\sigma}$.  The invariance 
of the inner product implies conservation of probability [20,24].  

\medskip
\noindent{\bf 3.  The unitarity constraints}

%--------|---------|---------|---------|---------|---------|---------|
\noindent The condition that the global evolution of a QCA be unitary
places strict constraints on the local rule $f$.  These constraints
can be expressed directly in terms of the amplitude vectors:

%--------|---------|---------|---------|---------|---------|---------|
\noindent\Theorem\ 3.1.  {\sl $F$ is unitary iff for all admissible 
configurations $\sigma'$ and $\sigma''$,
$$
\prod_{x \in L} \llangle\sigma''_{x+E} | \sigma'_{x+E}\rrangle = 
 \delta_{\sigma''\sigma'},                                  \eqno(3.1)
$$
and, in the infinite case, $F$ is surjective.}

%--------|---------|---------|---------|---------|---------|---------|
\noindent\Proof.  $F$ is unitary iff it preserves inner products and 
is surjective.  For finite dimensional $F$ the former condition 
implies the latter, as well as that $F^{-1} = F^{\dagger}$.  In the 
infinite case both conditions are necessary and sufficient to reach 
this conclusion [25].  The condition that $F$ preserves inner
products is equivalent to $I = F^{\dagger} F$; in components this
becomes:
$$
\eqalignno{
\delta_{\sigma''\sigma'} 
 &= \sum_{\sigma \in A} 
    (F^{\dagger})_{\sigma''\sigma} F_{\sigma\sigma'}
  = \sum_{\sigma \in A} 
    \overline F_{\sigma\sigma''} F_{\sigma\sigma'}                 \cr
 &= \sum_{\sigma \in A} 
    \overline{\prod_{x \in L} 
              f(\sigma^{\vphantom\prime}_{x\vphantom{E}}|
                \sigma''_{x+E})}
              \prod_{y \in L} 
              f(\sigma^{\vphantom\prime}_{y\vphantom{E}}|
                \sigma'_{y+E})                                     \cr
 &= \sum_{\sigma \in A} 
    \prod_{x \in L} 
    \overline{f(\sigma^{\vphantom\prime}_{x\vphantom{E}}|
                \sigma''_{x+E})}
              f(\sigma^{\vphantom\prime}_{x\vphantom{E}}|
                \sigma'_{x+E})                                     \cr
 &= \cdots\!\!\sum_{\sigma_{x_i} \in Q}\!\!\cdots\;
    \prod_{x \in L} 
    \overline{f(\sigma^{\vphantom\prime}_{x\vphantom{E}}|
                \sigma''_{x+E})}
              f(\sigma^{\vphantom\prime}_{x\vphantom{E}}|
                \sigma'_{x+E}),                                    \cr
\noalign{\hbox{(where the ellipses denote sums for each $x_i \in L$ 
and by admissibility condition ({\it iii\/}), only}
\vskip-2pt}
\noalign{\hbox{$\sigma \in A$ make nonzero contributions)}
\vskip2pt}
 &= \prod_{x \in L} \sum_{\sigma_x \in Q}
    \overline{f(\sigma^{\vphantom\prime}_{x\vphantom{E}}|
                \sigma''_{x+E})}
              f(\sigma^{\vphantom\prime}_{x\vphantom{E}}|
                \sigma'_{x+E})                                     \cr
 &= \prod_{x \in L} \llangle\sigma''_{x+E}|\sigma'_{x+E}\rrangle,  \cr
}
$$
which is, of course, no more than the familiar condition for unitarity
we described in [8] from the perspective of the sum-over-histories 
approach to quantum mechanics [26], written in terms of the transition
amplitudes $(2.2b)$.                                   \hfill\endproof

%--------|---------|---------|---------|---------|---------|---------|
{\it A priori\/} Theorem 3.1 tells us only that inner product 
invariance requires a constraint for each pair of admissible 
configurations.  To be able to verify unitarity for arbitrarily large 
or infinite lattices given a local rule we must reduce the number to 
something independent of the cardinality of $L$.  We do so in 
Subsections 3.1 and 3.2; then in Subsection 3.3 we consider the 
surjectivity condition which must also be satisfied for infinite 
lattices.  Our results in this section are similar to those obtained
independently by D\"urr, Thanh and Santha [18,19].

\noindent{\it 3.1.  The normalization constraints $\ldots$}

%--------|---------|---------|---------|---------|---------|---------|
There is one diagonal constraint in (3.1) for each admissible 
configuration:
$$
1 = 
 \prod_{x \in L} \llangle \sigma_{x+E}|\sigma_{x+E}\rrangle.\eqno(3.2)
$$
That is, for each admissible configuration, the product of the lengths 
of the amplitude vectors of its $E$-subconfigurations must be 1.  So 
to understand the structure of this set of constraints, we must 
understand which sets of $E$-subconfigurations can occur together.
Since our configurations are one dimensional they may be constructed
by appending (and prepending) one element of $Q$ after another; each
successive element is the right (left) most component of a 
corresponding $E$-subconfiguration which is determined by the new
element together with the previous $k-1$ elements.  

%--------|---------|---------|---------|---------|---------|---------|
This construction may be realized by a {\sl directed graph\/} (known 
as a de~Bruijn graph [27]) which is weighted [28]:  Let $\Gone$ be the 
$(q,q)$-valent directed graph with vertices labelled by the elements 
of $Q^{k-1}$ and directed edges labelled by $i_k \in Q$ connecting 
vertices labelled $i_1i_2\ldots i_{k-1}$ to vertices labelled 
$i_2\ldots i_{k-1}i_k$.  (See Figures 1 and 3 in Section 4 for 
examples.)   A {\sl path\/} of length $l$ in $\Gone$ is a sequence of 
$l$ directed edges in the graph, each of which (but the first) is 
directed away from the vertex to which the previous edge is directed.  
Paths of length $l$ correspond bijectively to subconfigurations of 
length $l + k - 1$; each edge corresponds to an $E$-subconfiguration.%
\sfootnote*{Equivalently, $\Gone$ is a {\sl finite state 
automaton\/} (FSA; see, \eg, [29]) with states corresponding to length 
$k-1$ subconfigurations and transitions to single elements of $Q$.  
Such FSAs model circuits with memory, like convolutional coders [30].  
When the outputs of an FSA are (state,transition) pairs it is called a 
{\sl Mealy machine\/} [31]; here the outputs are 
$E$-subconfigurations.}
A {\sl cycle\/} in a directed graph is a closed path which passes
through no vertex more than once.

%--------|---------|---------|---------|---------|---------|---------|
Assign the weight $\llangle i_1i_2\ldots i_k|i_1i_2\ldots i_k\rrangle$ 
to each edge labelled $i_k$ leaving a vertex labelled 
$i_1i_2\ldots i_{k-1}$ and define the weight of a path in the directed 
graph to be the product of the weights of its edges.  Then the weight 
of the path corresponding to configuration $\sigma$ is given by the 
right hand side of (3.2).

%--------|---------|---------|---------|---------|---------|---------|
\noindent\Lemma\ 3.2.  {\sl If (3.2) holds for every admissible
configuration then the weight of every cycle in $\Gone$ is 1.}

%--------|---------|---------|---------|---------|---------|---------|
\noindent\Proof.  Since any cycle corresponds to an admissible 
configuration $\sigma$ of a periodic CA for some $N$, (3.2) must hold 
for $\sigma$, proving the statement in the periodic case.  In the 
infinite case, consider any cycle with edges labelled $i_1\ldots i_l$ 
and a bi-infinite path in $\Gone$ containing the cycle which 
corresponds to an admissible configuration $\sigma$.  $\sigma$ must
have a subconfiguration corresponding to the cycle starting after some 
$x_0 \in \Z$:  
$\sigma_{x_0 + j} = i_j$ for $1 \le j \le l$ and 
$\sigma_{x_0 + l + j} = i_j$ for 
$0 \le j \le k-1$.  Define a new configuration $\sigma'$ by
$$
\sigma'_x = \cases{\sigma_x     & if $x \le x_0$;                  \cr
                   \sigma_{x + l} & if $x > x_0$.                  \cr
                   }
$$
$\Omega^{\pm}(\sigma') = \Omega^{\pm}(\sigma)$, so $\sigma'$ is 
admissible by Lemma 2.1.  Furthermore, the product in (3.2) will be 
identical for $\sigma$ and $\sigma'$ but for an extra product in the 
former of the weights of edges in the cycle.  Since (3.2) holds for 
both $\sigma$ and $\sigma'$, the weight of the cycle must be 1.
                                                       \hfill\endproof

%--------|---------|---------|---------|---------|---------|---------|
\noindent\Theorem\ 3.3.  {\sl For a periodic CA, (3.2) holds for all
admissible configurations iff the weight of every cycle in $\Gone$ 
is 1.}

%--------|---------|---------|---------|---------|---------|---------|
\noindent\Proof.  Lemma 3.2 is the `only if' part of this statement.
The `if' part is almost a tautology since the only configurations in a 
periodic CA correspond to collections of cycles, so if each has weight 
1, then (3.2) holds for all admissible configurations. \hfill\endproof

%--------|---------|---------|---------|---------|---------|---------|
The analogous result in the infinite case is somewhat more 
complicated because the admissible configurations are different.  Let
us examine them more closely:

%--------|---------|---------|---------|---------|---------|---------|
\noindent\Lemma\ 3.4.  {\sl Let $\sigma$ be any configuration in an 
infinite CA.  Then $\Omega^{\pm}(\sigma)$ consists of 
$E$-subconfigurations corresponding to the edges in a collection of 
cycles in $\Gone$.}
\eject

%--------|---------|---------|---------|---------|---------|---------|
\noindent\Proof.  We must show that for any $E$-subconfiguration 
$\lambda \in \Omega^+(\sigma)$ there is some cycle in $\Gone$ 
containing the edge corresponding to $\lambda$ such that the 
$E$-subconfiguration corresponding to each edge in the cycle is also 
in $\Omega^+(\sigma)$.  By definition, $\lambda \in \Omega^+(\sigma)$ 
only if the corresponding edge in $G_1(Q,k)$ is traversed infinitely 
many times by the infinite forward path corresponding to $\sigma$.  To 
return to this edge the infinite forward path must contain a cycle; 
to return to it infinitely many times it must contain at least one of 
the only finitely many cycles infinitely many times.  Each edge in 
that cycle is therefore traversed infinitely many times by the 
infinite forward path corresponding to $\sigma$ and hence the 
corresponding $E$-subconfigurations are in $\Omega^+(\sigma)$.  The 
analogous proof with `forward' replaced by `backward' proves the 
statement for $\Omega^-(\sigma)$.                      \hfill\endproof

%--------|---------|---------|---------|---------|---------|---------|
Applying Lemma 2.1, we have the immediate:

%--------|---------|---------|---------|---------|---------|---------|
\noindent\Corollary\ 3.5.  {\sl An infinite CA has admissible 
configurations only if there is a subset $D_{\! f} \subset B_{\! f}$ 
of local configurations corresponding to the edges in a collection of 
cycles of $\Gone$.}

%--------|---------|---------|---------|---------|---------|---------|
Let $\Done$ be the subgraph of $\Gone$ consisting of these
cycles.  A maximal $D_{\! f}$ which is closed under the evolution will
be referred to as the {\sl deterministic sector\/} of a QCA because of 
the following result:

%--------|---------|---------|---------|---------|---------|---------|
\noindent\Theorem\ 3.6.  {\sl If $F$ is unitary then when restricted 
to local configurations in $D_{\! f}$, the local rule is 
deterministic.  If $D_{\! f}$ is closed under the evolution then the 
set of configurations defined by
$A := \{\sigma \mid \Omega^{\pm}(\sigma) \subset D_{\! f}\}$ is 
admissible.}

%--------|---------|---------|---------|---------|---------|---------|
\noindent\Proof.  By Corollary 3.5, each 
$\lambda \in D_{\! f} \subset B_{\! f}$ corresponds to an edge in a 
$\Gone$ cycle, each edge of which corresponds to a local 
configuration $\lambda_i \in D_{\! f}$, $1 \le i \le l$.  By Theorem
3.1 and Lemma 3.2, if $F$ is unitary,
$$
1 = \prod_{i=1}^l \llangle\lambda_i|\lambda_i\rrangle.      \eqno(3.3)
$$
When we defined $B_{\! f}$ we observed that for each 
$\lambda \in B_{\! f}$, $|\lambda\rrangle$ has norm at least 1; now 
(3.3) implies that each $\lambda_i$ in this cycle, and hence every 
local configuration in $D_{\! f}$, has an amplitude vector with norm 
exactly 1.  Since each amplitude vector has at least one component 
equal to 1, all the remaining components must be 0; thus the local 
rule is deterministic.

%--------|---------|---------|---------|---------|---------|---------|
Since $F$ restricted to local configurations in $D_{\! f}$ is 
deterministic, for any $\sigma,\sigma' \in A$, either the ends of
$\sigma$ evolve to those of $\sigma'$, in which case 
$F_{\sigma'\sigma}$ is well defined; or they do not, in which case
$F_{\sigma'\sigma} = 0$, which is still well defined.  This verifies
admissibility condition ({\it i\/}).  Furthermore, $D_{\! f}$ is 
closed under the evolution means that for every path of length $k$ in 
$D_{\! f}$, corresponding to a subconfiguration $i_1\ldots i_{2k-1}$, 
the unique local configuration $i'_1\ldots i'_k$ defined by 
$f(i'_j|i^{\vphantom\prime}_j\ldots i^{\vphantom\prime}_{j+k-1}) = 1$,
$1 \le j \le k$, corresponds to an edge in $D_{\! f}$.  If $F$ is 
unitary then it is norm preserving, so for any $\sigma \in A$, there 
must be some $\sigma' \in \Z^Q$ with $F_{\sigma'\sigma} \not= 0$.  
Since $D_{\! f}$ is closed under the evolution, 
$\Omega^{\pm}(\sigma') \subset D_{\! f}$; this verifies admissibility 
conditions ({\it ii\/}) and, since $D_{\! f}$ is maximal, 
({\it iii\/}).                                         \hfill\endproof

%--------|---------|---------|---------|---------|---------|---------|
Since we only consider infinite CAs with admissible configurations, we
will assume henceforth that each has a nonempty deterministic sector
which defines the set of admissible configurations as in Theorem 3.6.
The remarkable consequence of this theorem is that an infinite QCA 
must be {\sl asymptotically deterministic\/}:  the only admissible 
configurations are those which evolve deterministically outside some
finite domain.  Indeterministic quantum evolution can only occur for 
a finite subconfiguration interpolating between the ends of the
configuration.  Thus the analogue of Theorem 3.3 for infinite CAs is:

%--------|---------|---------|---------|---------|---------|---------|
\noindent\Theorem\ 3.7.  {\sl For an infinite CA, (3.2) holds for all
admissible configurations iff the weight of every cycle in $\Gone$ 
is 1 and the weight of every acyclic path in $\Gone$ terminating at 
vertices in $\Done$ is 1.}

%--------|---------|---------|---------|---------|---------|---------|
\noindent\Proof.  By Lemma 3.2, if (3.2) holds for all admissible 
configurations then the weight of every cycle in $\Gone$ is 1.  
Consider any acyclic path connecting cycles in $\Done$.  By Theorem 
3.6, there is an admissible configuration corresponding to a 
bi-infinite path contained in $\Done$ except for a finite segment 
along the chosen acyclic path.  If (3.2) holds for all admissible
configurations the weight of the whole path is 1, as are the weights 
of the two ends in $\Done$; hence the weight of the acyclic path is 
also 1.  Conversely, by definition and by Lemma 3.4, the only 
admissible configurations are those with ends corresponding to 
collections of cycles, connected by a finite subconfiguration which, 
after removing some finite number of subconfigurations corresponding 
also to cycles in $\Gone$, is an acyclic path terminating at vertices
in $\Done$.  If all the cycles and the acyclic path have weight 1 then 
(3.2) holds.                                           \hfill\endproof

\noindent{\it 3.2.  $\ldots$ and the orthogonality constraints}
\nobreak

\nobreak
%--------|---------|---------|---------|---------|---------|---------|
There is one off-diagonal constraint in (3.1) for each pair of 
distinct admissible configurations:
$$
0 = \prod_{x\in L} \llangle\sigma''_{x+E}|\sigma'_{x+E}\rrangle. 
                                                            \eqno(3.4)
$$
By (3.2), for each $x$ such that $\sigma''_{x+E} = \sigma'_{x+E}$, the
corresponding factor in (3.4) is nonzero.  Thus at least one of the 
pairs of {\sl mismatched\/} $E$-subconfigurations 
$\sigma''_{x+E} \not= \sigma'_{x+E}$ must contribute a factor of 0 in 
order for the product in (3.4) to vanish.  To understand this set of
orthogonality constraints, therefore, we must understand which sets of
pairs of mismatched $E$-subconfigurations can occur together.  We
construct a new weighted directed graph which generates these sets:

%--------|---------|---------|---------|---------|---------|---------|
Let $\Gtwo$ be the $(q^2,q^2)$-directed graph with vertices 
labelled by the elements of $Q^{k-1} \times Q^{k-1}$ and directed 
edges labelled by $(i''_k,i'_k) \in Q^2$ connecting vertices labelled 
$(i''_1i''_2 \ldots i''_{k-1},i'_1i'_2 \ldots i'_{k-1})$ to vertices
labelled
$(i''_2 \ldots i''_{k-1}i''_k,i'_2 \ldots i'_{k-1}i'_k)$.  Note that 
there is a subgraph of $\Gtwo$ isomorphic to $\Gone$, namely
those vertices and edges with both components of these labels 
identical; so we write $\Gone \subset \Gtwo$.  (See Figure 2 in 
Section 4 for an example.)

%--------|---------|---------|---------|---------|---------|---------|
Paths in $\Gtwo$ correspond bijectively to {\sl pairs\/} of 
subconfigurations; each edge corresponds to a pair of 
$E$-subconfigurations.  We assign the weight 
$\llangle i''_1i''_2\ldots i''_k|i'_1i'_2\ldots i'_k\rrangle$ to each
edge labelled $(i''_k,i'_k)$ leaving a vertex labelled 
$(i''_1i''_2 \ldots i''_{k-1},i'_1i'_2 \ldots i'_{k-1})$ so that the
weight of the path corresponding to a pair of configurations 
$\sigma'$ and $\sigma''$ is given by the right hand side of (3.4).  
Two configurations $\sigma'$ and $\sigma''$ are distinct only if they
have at least one pair of mismatched $E$-subconfigurations, so a path
in $\Gtwo$ corresponds to them only if it intersects 
$M(Q,k) := \Gtwo \setminus \Gone$.

%--------|---------|---------|---------|---------|---------|---------|
\noindent\Lemma\ 3.8.  {\sl If $F$ is unitary then the weight of every
acyclic path in $M(Q,k)$ terminating at vertices of $\Gone$ is 0.}

%--------|---------|---------|---------|---------|---------|---------|
\noindent\Proof.  Any acyclic path in $M(Q,k)$ terminating at vertices
of $\Gone$ can be extended, using only edges in $\Gone$, to a path in
$\Gtwo$ corresponding to a pair of distinct admissible configurations
$\sigma'$ and $\sigma''$.  If $F$ is unitary, however, Theorem 3.1 
implies condition (3.2), which precludes the weight of any edge in 
$\Gone$ from vanishing.  Thus (3.4) can only be satisfied for 
$\sigma'$ and $\sigma''$ if the weight of the portion of the path in
$M(Q,k)$ vanishes. $\phantom{some space}$              \hfill\endproof

%--------|---------|---------|---------|---------|---------|---------|
\noindent\Theorem\ 3.9.  {\sl For a periodic CA, $F$ is unitary iff
all of the following are true:
\parskip=0pt\setbox1=\hbox{({\it iii})\enspace}\parindent=\wd1
\medskip
\item{  ({\it i})}The weight of every cycle in $\Gone$ is 1.
\item{ ({\it ii})}The weight of every cycle in $M(Q,k)$ is 0.
\item{({\it iii})}The weight of every acyclic path in $M(Q,k)$ 
                  terminating at vertices in $\Gone$ is 0.

}
\parskip=10pt\parindent=20pt
%--------|---------|---------|---------|---------|---------|---------|
\noindent\Proof.  If $F$ is unitary then Theorem 3.1 and Lemma 3.2 
imply ({\it i\/}) while Theorem 3.1 and Lemma 3.8 imply ({\it iii\/}).  
Any cycle in $M(Q,k)$ corresponds to a pair of distinct admissible 
configurations $\sigma'$ and $\sigma''$ of a periodic CA for some $N$.  
Theorem 3.1 implies (3.4) must hold for $\sigma'$ and $\sigma''$; this 
implies ({\it ii\/}).

%--------|---------|---------|---------|---------|---------|---------|
Conversely, by Theorem 3.3, ({\it i\/}) implies (3.2) holds for all 
admissible configurations in a periodic CA.  Furthermore, pairs of 
distinct admissible configurations in a periodic CA correspond to 
collections of cycles in $\Gtwo$, at least one of which intersects 
$M(Q,k)$.  Thus (3.4) holds if each cycle which intersects $M(Q,k)$ 
has weight 0; ({\it ii\/}) and ({\it iii\/}) imply that this is the 
case.  By Theorem 3.1, (3.2) and (3.4) imply $F$ is unitary.  
                                                       \hfill\endproof

%--------|---------|---------|---------|---------|---------|---------|
Before stating the analogous result in the infinite case, we define
$\Dtwo$ to be the subgraph of $\Gtwo$ consisting of the edges 
corresponding to pairs of local configurations each of which is in 
$D_{\! f}$.

%--------|---------|---------|---------|---------|---------|---------|
\noindent\Theorem\ 3.10.  {\sl For an infinite CA, $F$ is unitary iff
all of the following are true:
\parskip=0pt\setbox1=\hbox{({\it iii})\enspace}\parindent=\wd1
\medskip
\item{  ({\it i})}The weight of every cycle in $\Gone$ is 1.
\item{ ({\it ii})}The weight of every acyclic path in $\Gone$ 
                  terminating at vertices in $\Done$ is 1.
\item{({\it iii})}The weight of every acyclic path in $M(Q,k)$ 
                  terminating at vertices in $\Gone$ is 0.
\item{ ({\it iv})}The weight of every cycle in $\Dtwo \cap M(Q,k)$ is 
                  0. 
\item{  ({\it v})}$F$ is surjective.

}
\parskip=10pt\parindent=20pt
%--------|---------|---------|---------|---------|---------|---------|
\noindent\Proof.  If $F$ is unitary then Theorem 3.1 and Theorem 3.7
imply ({\it i\/}), ({\it ii\/}) and ({\it v\/}),  while Theorem 3.1 and 
Lemma 3.8 imply ({\it iii\/}).  Any cycle in $\Dtwo \cap M(Q,k)$ can 
be repeated infinitely and then corresponds to a pair of distinct 
admissible configurations $\sigma'$ and $\sigma''$ (since the ends of 
each are contained in $D_{\! f}$).  Theorem 3.1 implies (3.4) must 
hold for $\sigma'$ and $\sigma''$; this implies ({\it iv\/}).

%--------|---------|---------|---------|---------|---------|---------|
Conversely, by Theorem 3.7, ({\it i\/}) and ({\it ii\/}) imply (3.2) 
holds for all admissible configurations in an infinite CA.  
Furthermore, pairs of distinct admissible configurations in an 
infinite CA correspond to bi-infinite paths in $\Gtwo$ which consist 
of cycles in $\Dtwo$ at both ends.  If any one of these cycles lies in 
$\Dtwo \cap M(Q,k)$ then ({\it iv\/}) implies (3.4) holds.  If none 
of the cycles contains only edges corresponding to mismatched 
$E$-subconfigurations then there must be an acyclic path in $M(Q,k)$ 
terminating at vertices in $\Gone$.  In this case ({\it iii\/}) 
implies (3.4) holds.  By Theorem 3.1, ({\it v\/}), (3.2) and (3.4) 
imply $F$ is unitary.                                  \hfill\endproof

\noindent{\it 3.3.  Surjectivity}
\nobreak

\nobreak
%--------|---------|---------|---------|---------|---------|---------|
To complete the program of determining a collection of constraints on
the amplitude vectors which is equivalent to unitarity of the global
evolution we must show that for infinite CAs, condition ({\it v\/}) of
Theorem 3.10---surjectivity---is equivalent to some collection of such
constraints.  

%--------|---------|---------|---------|---------|---------|---------|
By Theorem 3.6 each $\sigma \in A$ has a minimal length 
{\sl interior\/} subconfiguration 
$I(\sigma) = \sigma_{x_0+1}\ldots\sigma_{x_0+n}$, defined by the 
condition that if $j < x_0+1$ or $j > x_0+n+k-1$ then 
$\sigma_{j-k+1}\ldots\sigma_j \in \Omega^{\pm}(\sigma)$, respectively.  
We will refer to the semi-infinite subconfigurations 
$\ldots\sigma_{x_0}$ and $\sigma_{x_0+n+1}\ldots$ as the 
{\sl deterministic ends\/} of $\sigma$.  For example, if 
$D_{\! f} = \{000,111\}$ ($k = 3$) then a possible admissible 
configuration is $\sigma = \ldots 0\sigma_1\ldots\sigma_n 1\ldots$ 
and $I(\sigma) = \sigma_1\ldots\sigma_n$, provided 
$\sigma_1 \not=0$ and $\sigma_n \not= 1$.  Let 
$A_n := \{\sigma \in A \mid |I(\sigma)| = n\}$.  Then $A = \cup A_n$, 
$0\le n < \infty$.  We will find a necessary and sufficient set of
constraints on $f$ such that $F$ maps onto $A_n$ for each nonnegative 
integer $n$; since $A$ is an orthonormal basis for $H$ this is 
equivalent to $F$ mapping onto $H$.

%--------|---------|---------|---------|---------|---------|---------|
We begin by observing that $F$ already maps onto the {\sl completely 
deterministic configurations\/} $A_0$.  More precisely,

%--------|---------|---------|---------|---------|---------|---------|
\noindent\Lemma\ 3.11.  {\sl For an infinite CA, if $F$ is norm 
preserving (\ie, if conditions ({\it i})---({\it iv}) of Theorem 3.10
hold) then $F : A_0 \to A_0$ is surjective.}

%--------|---------|---------|---------|---------|---------|---------|
\noindent\Proof.  If $F$ is norm preserving it is injective on 
$H \supset A_0$.  By definition, all $E$-subconfigura\-tions of $A_0$ 
are in $D_{\! f}$, so $F$ is deterministic on $A_0$.%
\sfootnote*{Notice that neither Moore and Myhill's Garden of Eden 
Theorem [32] nor Hedlund's result that injectivity implies 
surjectivity for endomorphisms of shift dynamical systems [33] may be
applied to conclude that $F$ is surjective on $A_0$ since $A$, and
hence $A_0$, is only a proper subset of the set of all possible 
configurations.}
Every configuration $\sigma \in A_0$ corresponds to a sequence of 
cycles in $\Done$ since $I(\sigma) = \emptyset$.  But injectivity 
implies surjectivity for $F$ restricted to cycles in $\Done$:  Each 
cycle of length $m$ must map to a closed path of length $m$ which, by
injectivity, must also be a cycle.  There are only a finite number of 
length $m$ cycles and by injectivity $F$ maps no two to the same one;
hence $F$ is surjective on cycles in $\Done$.          \hfill\endproof

%--------|---------|---------|---------|---------|---------|---------|
This result has an immediate corollary once we make some suitable
definitions:

%--------|---------|---------|---------|---------|---------|---------|
\noindent\Corollary\ 3.12.  {\sl For an infinite CA, if $F$ is norm
preserving then the restriction of $F$ to semi-infinite completely 
deterministic admissible subconfigurations is surjective.}

%--------|---------|---------|---------|---------|---------|---------|
To define the restriction of $F$ on subconfigurations it will be 
convenient in the rest of this subsection to take neighborhoods of 
size $k$ to be defined by $E := \{0,\ldots,k-1\}$.  Any other 
connected local neighborhood of the same size is equivalent to this 
choice after a translation, so there is no loss of generality.  Having 
made this choice for $E$, we may define the restriction of $F$ to any 
subconfiguration $\sigma_a\ldots\sigma_b$ of any admissible 
configuration by
$$
F|\sigma_a\ldots\sigma_b\rangle
 := \sum_{\sigma'_a\ldots\sigma'_{b-k+1}}
     |\sigma'_a\ldots\sigma'_{b-k+1}\rangle                 
     \prod_{i=a}^{b-k+1}
      f(\sigma'_i|
        \sigma^{\vphantom\prime}_i\ldots
        \sigma^{\vphantom\prime}_{i+k-1}),
                                                            \eqno(3.5)
$$
which is well defined even when either $a = -\infty$ or $b = \infty$
since $\sigma_a\ldots\sigma_b$ is a subconfiguration of an admissible 
configuration.  ((3.5) is equivalent to ($2.2a,b$) when $a = -\infty,
b = \infty$.)  Now we can prove Corollary 3.12:

%--------|---------|---------|---------|---------|---------|---------|
\noindent\Proof.  By definition, a semi-infinite completely 
deterministic admissible subconfiguration is a deterministic end of at 
least one configuration in $A_0$.  By Lemma 3.11, $F$ is surjective on 
$A_0$; whence the restriction of $F$ defined by (3.5) to 
subconfigurations of configurations in $A_0$ is also surjective.
                                                       \hfill\endproof

%--------|---------|---------|---------|---------|---------|---------|
This means that $F$ restricted to the deterministic ends of admissible
configurations is surjective and indicates that we may concentrate 
just on interior subconfigurations.  For 
$\lambda = \lambda_0\ldots\lambda_{k-1}$, 
$\rho = \rho_0\ldots\rho_{k-1} \in D_{\! f}$ define
$$
\eqalign{
A_n^{(\rho)} 
 &:= \{i_1\ldots i_n\rho_0\ldots\rho_{k-2} \mid 
       i_j\in Q, 1 \le j \le n\}                                   \cr
A_n^{(\lambda,\rho)}
 &:= \{\lambda_1\ldots\lambda_{k-1}i_1\ldots i_n\rho_0\ldots\rho_{k-2} 
       \mid i_j\in Q, 1 \le j \le n\}.                             \cr
}
$$
Considering an element of $A_n^{(\lambda,\rho)}$ to be a 
subconfiguration of an admissible configuration we may use (3.5) to 
define the action of $F$ on $A_n^{(\lambda,\rho)}$.

%--------|---------|---------|---------|---------|---------|---------|
\noindent\Theorem\ 3.13.  {\sl For an infinite CA with $F$ norm 
preserving, $F$ maps onto $A_n$ for $0 < n \in \Z$ iff for all 
$\lambda,\rho' \in D_{\! f}$, its restriction 
$F_n^{(\lambda,\rho')} : A_n^{(\lambda,\rho)} \to A_n^{(\rho')}$ is 
surjective when $f(\rho'_{k-1}|\rho) = 1$.}

%--------|---------|---------|---------|---------|---------|---------|
\noindent\Proof.  Each configuration $\sigma' \in A_n$ has an 
interior subconfiguration of length $n$ extended by deterministic 
ends:  
$\sigma' = \ldots\lambda'_{k-1}i'_1\ldots i'_n\rho'_0\ldots$ where
$\sigma'_{x_0} = i'_1$, say, and 
$\lambda'_{j-k+1}\ldots\lambda'_j \in D_{\! f}$ for $j \le k-1$,
$\rho'_j\ldots\rho'_{j+k-1} \in D_{\! f}$ for $j \ge 0$.  $F$ maps 
onto $A_n$ iff for all $\sigma' \in A_n$,
$$
|\sigma'\rangle = F\sum c_m|\sigma^m\rangle                 \eqno(3.6)
$$
for some element $\sum c_m|\sigma^m\rangle \in H$.  By Corollary 3.12,
there exist semi-infinite deterministic configurations 
$\ldots\lambda_{k-1}$ and $\rho_0\ldots$ such that 
$\langle\sigma'|F|\sigma^m\rangle \not= 0$ only for configurations of
the form 
$\sigma^m = \ldots\lambda^{\vphantom\prime}_0
            \ldots\lambda^{\vphantom\prime}_{k-1}
            i^m_1\ldots i^m_n\rho^{\vphantom\prime}_0\ldots$, 
where $\sigma^{\vphantom\prime}_{x_0} = \lambda^{\vphantom\prime}_1$ 
and
$$
\eqalign{
F|\ldots\lambda^{\vphantom\prime}_0
  \ldots\lambda^{\vphantom\prime}_{k-1}\rangle 
 &= |\ldots\lambda'_{k-1}\rangle                                   \cr
F|\rho^{\vphantom\prime}_0\ldots\rangle 
 &= |\rho'_0\ldots\rangle.                                         \cr
}                                                           \eqno(3.7)
$$
For these configurations (3.5) implies that
$$
\langle\sigma'|F|\sigma^m\rangle 
 = \langle i'_1\ldots i'_n\rho'_0\ldots\rho'_{k-2}|F|
    \lambda^{\vphantom\prime}_1\ldots\lambda^{\vphantom\prime}_{k-1}
    i^m_1\ldots i^m_n
    \rho^{\vphantom\prime}_0\ldots\rho^{\vphantom\prime}_{k-2}\rangle.
                                                            \eqno(3.8)
$$
Combining (3.6), (3.7) and (3.8) we find
$$
|i'_1\ldots i'_n\rho'_0\ldots\rho'_{k-2}\rangle =
 F\sum c_m|\lambda^{\vphantom\prime}_1\ldots
           \lambda^{\vphantom\prime}_{k-1}
           i^m_1\ldots i^m_n
           \rho^{\vphantom\prime}_0\ldots
           \rho^{\vphantom\prime}_{k-2}\rangle,
                                                            \eqno(3.9)
$$
which is exactly the statement that $F_n^{(\lambda,\rho')}$ is 
surjective.  Since 
$\lambda' := \lambda'_0\ldots\lambda'_{k-1} \in D_{\! f}$ is arbitrary
and $f(\lambda'_{k-1}|\lambda) = 1$ (from (3.7)) defines a unique
$\lambda := \lambda_0\ldots\lambda_{k-1} \in D_{\! f}$ by
Corollary 3.12, $\lambda$ is an arbitrary deterministic local 
configuration.  Furthermore, (3.7) implies $f(\rho'_{k-1}|\rho) = 1$ 
for $\rho := \rho_0\ldots\rho_{k-1}$; this completes the ``only if''
direction of the proof.  Conversely, if for all 
$\lambda,\rho' \in D_{\! f}$ with $f(\rho'_{k-1}|\rho) = 1$, (3.9) 
holds for some $c_m \in \C$, then using Corollary 3.12 we can 
construct arbitrary deterministic ends satisfying (3.7) and conclude 
that (3.6) holds for all $\sigma' \in A_n$.            \hfill\endproof

%--------|---------|---------|---------|---------|---------|---------|
With Lemma 3.11 and Theorem 3.13 we have reduced the problem of the
surjectivity of the infinite dimensional map $F$ to the equivalent
problem of the surjectivity of the finite dimensional maps $F_n$ for 
all $1 \le n \in \Z$ (we will supress the superscripts 
$(\lambda,\rho')$ in the following).  But $F_n$ is onto iff 
$\det F_n \not= 0$.  To understand these conditions we must explicate
the structure of $F_n$.  Notice first that as an immediate consequence 
of definition (3.5) each column of $F_n$ has a common factor.  More 
precisely,

%--------|---------|---------|---------|---------|---------|---------|
\noindent\Lemma\ 3.14.  {\sl For $\alpha,\alpha' \in Q^n$, 
$$
(F_n)_{\alpha',\alpha} 
 = (\tilde F_n)_{\alpha',\alpha} \cdot
   \cases{\langle
           \rho'_0\ldots\rho'_{k-2}|F|
           \alpha^{\vphantom\prime}_{n-k+2}\ldots
           \alpha^{\vphantom\prime}_n
           \rho^{\vphantom\prime}_0\ldots\rho^{\vphantom\prime}_{k-2}
          \rangle, & if\/ $n \ge k-1${\rm ;}                       \cr
          \langle
           \rho'_0\ldots\rho'_{k-2}|F|
           \lambda^{\vphantom\prime}_{n+1}\ldots
           \lambda^{\vphantom\prime}_{k-1}\alpha
           \rho^{\vphantom\prime}_0\ldots\rho^{\vphantom\prime}_{k-2}
          \rangle, & if\/ $0 \le n < k-1${\rm ;}                   \cr
         }                                                 \eqno(3.10)
$$
where $\tilde F$ is the\/} reduced transition matrix {\sl defined by
$$
(\tilde F_n)_{\alpha',\alpha} 
 := \langle \alpha'|F|\lambda_1\ldots\lambda_{k-1}\alpha \rangle.
                                                           \eqno(3.11)
$$
}

%--------|---------|---------|---------|---------|---------|---------|
\noindent These common factors can be pulled out of each column in the
determinant:

%--------|---------|---------|---------|---------|---------|---------|
\noindent\Corollary\ 3.15.  {\sl $\det F_n = c_n \det \tilde F_n$,
where
$$
c_n := \cases{\prod_{\gamma\in Q^{k-1}}
               \langle
                \rho'_0\ldots\rho'_{k-2}|F|
                \gamma
                \rho^{\vphantom\prime}_0\ldots
                \rho^{\vphantom\prime}_{k-2}
               \rangle^{q^{n-k+1}},  & if\/ $n \ge k-1${\rm ;}     \cr
              \prod_{\alpha\in Q^n}
               \langle
                \rho'_0\ldots\rho'_{k-2}|F|
                \lambda^{\vphantom\prime}_{n+1}\ldots
                \lambda^{\vphantom\prime}_{k-1}\alpha
                \rho^{\vphantom\prime}_0\ldots
                \rho^{\vphantom\prime}_{k-2}
               \rangle, & if\/ $0 \le n < k-1$.                    \cr
             }                                             \eqno(3.12)
$$
}

%--------|---------|---------|---------|---------|---------|---------|
\noindent\Proof.  When $n \ge k-1$, for each $\gamma \in Q^{k-1}$ 
there are $q^{n-k+1}$ columns in $F_n$ labelled by subconfigurations
having $\gamma$ as their last $k-1$ states.  According to Lemma 3.14,
the factor in (3.10) occurs in each.  When $0 \le n < k-1$, for each
$\alpha \in Q^n$ there is exactly one column in $F_n$ labelled by 
$\alpha$.  Again, by Lemma 3.14, this column contains the factor in 
(3.10).                                                \hfill\endproof

%--------|---------|---------|---------|---------|---------|---------|
Second, for $\gamma \in Q^{k-1}$, let 
$$
\Phi^{(\gamma)} := \pmatrix{|\gamma 0    \rrangle \cr
                            \vdots                \cr
                            |\gamma (q-1)\rrangle \cr
                           } \in M_q(\C).                  \eqno(3.13)
$$
Now we can find a recurrence relation (with initial condition 
$\tilde F_0 = 1$) for the reduced transition matrix:

%--------|---------|---------|---------|---------|---------|---------|
\noindent\Lemma\ 3.16.  {\sl $\tilde F_{n+1}$ can be expressed in 
terms of $\tilde F_n$ and the $\Phi^{(\gamma)}$ as
$$
(\tilde F_{n+1})_{\alpha' j,\alpha i}
 = (\tilde F_n)_{\alpha',\alpha} \cdot
   \cases{\Phi^{(\alpha_{n-k+2}\ldots\alpha_n)}_{ij},  
           & if\/ $n \ge k-1${\rm ;}                               \cr
          \Phi^{(\lambda_{n+1}\ldots\lambda_{k-1}\alpha)}_{ij},
           & if\/ $0 \le n < k-1${\rm ;}                           \cr
         }                                                 \eqno(3.14)
$$
where $\alpha,\alpha' \in Q^n$ and $i,j \in Q$.}

%--------|---------|---------|---------|---------|---------|---------|
\noindent\Proof.  Apply definition (3.11):
$$
\eqalign{
(\tilde F_{n+1})_{\alpha' j,\alpha i}
 &= \langle 
     \alpha' j|F|\lambda_1\ldots\lambda_{k-1}\alpha i
    \rangle                                                        \cr
 &= \langle 
     \alpha' |F|\lambda_1\ldots\lambda_{k-1}\alpha
    \rangle
    \cdot
    \cases{f(j|\alpha_{n-k+2}\ldots\alpha_n i),  
            & if\/ $n \ge k-1$;                                    \cr
           f(j|\lambda_{n+1}\ldots\lambda_{k-1}\alpha i),
            & if\/ $0 \le n < k-1$;                                \cr
          }
}
$$
where the second equality follows from definition (3.5).  Definition
(3.13) gives the result (3.14).                        \hfill\endproof

%--------|---------|---------|---------|---------|---------|---------|
Just as the common factors in Lemma 3.14 could be pulled out of the
determinant of $F_n$ in Corollary 3.15, so too can the 
{\sl determininants\/} of the common tensor factors $\Phi^{(\gamma)}$.
We need two simple properties of determinants:

%--------|---------|---------|---------|---------|---------|---------|
\itemitem{1.} For $i,j \in \{1,\ldots,m\}$, let $X_{ij} \in M_n(\C)$
be $m \times m$ matrices over $\C$.  Suppose $X_{i1} = x_i B$ with 
$x_i \in \C$ and $B \in M_n(\C)$.  Then
$$
\det\pmatrix{X_{11} & \cdots & X_{1m} \cr
             \vdots &        & \vdots \cr
             X_{m1} & \cdots & X_{mm} \cr
            }
 = \det B \det\pmatrix{ x_1 I & X_{12} & \cdots & X_{1m} \cr
                       \vdots & \vdots &        & \vdots \cr
                        x_m I & X_{m2} & \cdots & X_{mm} \cr
                      },                                   \eqno(3.15)
$$
for $I$ the $n\times n$ identity matrix.

%--------|---------|---------|---------|---------|---------|---------|
\itemitem{2.} Let $X \in M_m(\C)$.  Then
$$
\det(X \otimes I) = (\det X)^n                             \eqno(3.16)
$$
where $I$ is again the $n\times n$ identity matrix.

%--------|---------|---------|---------|---------|---------|---------|
\noindent\Corollary\ 3.17.  {\sl 
$\det \tilde F_{n+1} = d_{n+1} [\det \tilde F_n]^q$, where
$$
d_{n+1}
 := \cases{\prod_{\gamma\in Q^{k-1}}
           [\det\Phi^{(\gamma)}]^{q^{n-k+1}},
            & if\/ $n \ge k-1${\rm ;}                              \cr
           \prod_{\alpha\in Q^n}
           \det\Phi^{(\lambda_{n+1}\ldots\lambda_{k-1}\alpha)},
            & if\/ $0 \le n < k-1$.                                \cr
          }                                                \eqno(3.17)
$$
}

%--------|---------|---------|---------|---------|---------|---------|
\noindent\Proof.  By Lemma 3.16 the columns of $\tilde F_{n+1}$ come 
from the columns of $\tilde F_n$ tensored with the corresponding 
$\Phi^{(\gamma)}$.  Applying (3.15) we have:
$$
\det \tilde F_{n+1}
 = \det \tilde F_n \otimes I \cdot
   \cases{\prod_{\gamma\in Q^{k-1}}
           [\det\Phi^{(\gamma)}]^{q^{n-k+1}},
            & if $n \ge k-1$;                                      \cr
           \prod_{\alpha\in Q^n}
           \det\Phi^{(\lambda_{n+1}\ldots\lambda_{k-1}\alpha)},
            & if $0 \le n < k-1$                                   \cr
          }
$$
(where $I$ is the $q \times q$ identity matrix), by exactly the same
reasoning as in the proof of Corollary 3.15, but now using the result
of Lemma 3.16 for the common tensored matrices.  Applying (3.16) gives 
the result (3.17).                                     \hfill\endproof

%--------|---------|---------|---------|---------|---------|---------|
Putting these results together gives the Surjectivity Theorem 
summarizing the conditions that for all $1 \le n \in \Z$, 
$\det F_n \not= 0$:

%--------|---------|---------|---------|---------|---------|---------|
\noindent\Theorem\ 3.18.  {\sl An infinite CA with norm preserving $F$
and deterministic sector $D_{\! f}$ is surjective iff for all 
$\lambda,\rho,\rho' \in D_{\! f}$ such that $f(\rho'_{k-1}|\rho) = 1$,
none of the following vanish:
\parskip=0pt
\medskip
\itemitem{  ({\it i})}$\langle
                        \rho'_0\ldots\rho'_{k-2}|F|
                        \gamma
                        \rho^{\vphantom\prime}_0\ldots
                        \rho^{\vphantom\prime}_{k-2}
                       \rangle$
\itemitem{ ({\it ii})}$\det\Phi^{(\gamma)}$
\medskip
\noindent for any $\gamma \in Q^{k-1}$, $0 \le n < k-1$.
}
\parskip=10pt

%--------|---------|---------|---------|---------|---------|---------|
\noindent\Proof.  By Corollaries 3.15 and 3.17, $\det F_n \not= 0$ for
all $n \ge 0$ iff none of the $c_n$ defined in (3.12) nor the $d_n$ 
defined in (3.17) vanish.  This is
ensured by ({\it i\/}) and ({\it ii\/}), respectively, since
$$
\eqalignno{
\{|\lambda_{n+1}\ldots\lambda_{k-1}\alpha\rho_0\ldots\rho_{k-2}\rangle
  \mid \alpha\in Q^n, 0 \le n < k-1\} 
&\subset 
 \{|\gamma\rho_0\ldots\rho_{k-2}\rangle \mid \gamma \in Q^{k-1}\}  \cr
\noalign{\hbox{and}
\vskip2pt}
\{\Phi^{(\lambda_{n+1}\ldots\lambda_{k-1}\alpha)}
  \mid \alpha\in Q^n, 0 \le n < k-1\} 
&\subset 
 \{\Phi^{(\gamma)} \mid \gamma \in Q^{k-1}\}.                      \cr
}
$$
$\phantom{some space}$\hfill\endproof

%--------|---------|---------|---------|---------|---------|---------|
\noindent We will refer to the nonvanishing of the expressions in 
({\it i\/}) and ({\it ii\/}) as the {\sl surjectivity constraints}.  
Each is a closed condition, so as we will see in the next section, 
when they are not inconsistent with constraints 
({\it i\/})--({\it iv\/}) of Theorem 3.10, they do not reduce the 
dimension of the solution space.

\medskip
\noindent{\bf 4.  Solutions}

%--------|---------|---------|---------|---------|---------|---------|
\noindent Theorems 3.9 and 3.10, to which we will refer as the 
Unitarity Theorems, and Surjectivity Theorem 3.18, together with the 
finiteness of the graph $\Gtwo$, show that given a local rule $f$ we 
can determine, by checking only a finite number of conditions, whether 
the global evolution is unitary, \ie, whether $f$ defines a QCA.  This 
proves:

%--------|---------|---------|---------|---------|---------|---------|
\noindent\Theorem\ 4.1.  {\sl Unitarity is decidable for 1 dimensional
CAs.}

%--------|---------|---------|---------|---------|---------|---------|
\noindent As yet we have no proposed local rule $f$ for which to check 
unitarity.  In the next two subsections we write down the constraints 
resulting from the Unitarity Theorems for binary (\ie, $q = 2$) CAs in 
the two simplest cases and show that they can be solved to give 
multiparameter families of QCAs.

%--------|---------|---------|---------|---------|---------|---------|
Before doing so, however, it is useful to discuss some symmetries of 
QCAs.  First, note that the symmetry group of a one dimensional 
lattice $L$ has two generators:  $T$, translation by one and $P$, 
reflection in the origin/parity reversal.  $P$ acts on local 
configurations, sending $\lambda = i_1\ldots i_k$ to 
$P\lambda = i_k\ldots i_1$.  (To be precise, this is the case when $k$ 
is odd; when $k$ is even this transformation is $TP$ if the origin is 
taken to be at the $\lfloor k/2 \rfloor$ position in the local 
neighborhood.  It will cause no confusion, however, to denote both by 
$P$.)  $P$ may be defined to act on a local rule $f$ by:  
$(Pf)(i|\lambda) := f(i|P\lambda)$; hence 
$P|i_1\ldots i_k\rrangle := |i_k\ldots i_1\rrangle$.  We will refer to 
$Pf$ as the {\sl parity transform\/} of $f$.  Since $P^2f = f$, any 
local rule which is not {\sl symmetric}, \ie, invariant under this 
reflection, will pair with a distinct parity transformed local rule.

%--------|---------|---------|---------|---------|---------|---------|
Second, note that the symmetric group $S_q$ acts on the set of states
$Q$ of a QCA.  For $q = 2$ the symmetric group is generated by the
transposition $\tau$ which interchanges 0 and 1.%
\sfootnote*{From a more physical perspective, $\tau$ is analogous to
charge conjugation.}
$\tau$ acts on a local configuration $\lambda = i_1\ldots i_k$ to give 
$\tau\lambda = \tau i_1\ldots\tau i_k$, so it acts on a local rule in 
two ways:
$$
\eqalign{
(\tau_{\rm in}f)(i|\lambda) &:= f(i|\tau\lambda)                   \cr
(\tau_{\rm out}f)(i|\lambda) &:= f(\tau i|\lambda).                \cr
}
$$
These transformations will be most useful in the discussion of 
infinite, and hence asymptotically deterministic, QCAs.
\eject

\noindent{\it 4.1.  $k = 2$}
\nobreak

\nobreak
%--------|---------|---------|---------|---------|---------|---------|
\moveright\secondstart\vtop to 0pt{\hsize=\halfwidth
{}\vskip -\baselineskip
$$
\epsfxsize=\halfwidth\epsfbox{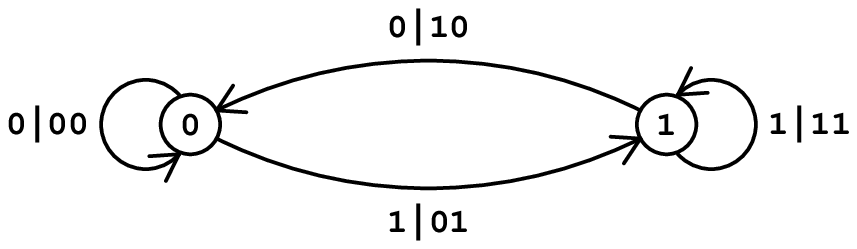}
$$
\vskip -\baselineskip
\eightpoint{%
\noindent{\bf Figure 1}.  $G_1$ for binary CAs with local 
neighborhoods of size 2.  Each vertex and edge is labelled, the latter
by the number to the left of the slash.  The weight of each edge is 
the squared norm of the amplitude vector of the local 
config\-u\-ration indicated to the right of the slash.
}}
\vskip -\baselineskip
\parshape=14
0pt \halfwidth
0pt \halfwidth
0pt \halfwidth
0pt \halfwidth
0pt \halfwidth
0pt \halfwidth
0pt \halfwidth
0pt \halfwidth
0pt \halfwidth
0pt \halfwidth
0pt \halfwidth
0pt \halfwidth
0pt \halfwidth
0pt \hsize
%--------|---------|---------|---------|---------|---------|---------|
{The simplest nontrivial binary CAs have local neighborhoods of size 
2.%
\sfootnote*{These are sometimes called `one-way' automata 
[15] since the local neighborhood of $x$ extends only to one side, 
although Hillman has pointed out that this terminology is somewhat 
misleading [34].}
Figure 1 shows $G_1$.  (The arguments $Q = \{0,1\}$ and $k = 2$ are
suppressed in this subsection.)  Here all the relevant paths 
can be identified by inspection.  More formally, define the transfer 
matrix $A_1$ for $G_1$ to be the $2 \times 2$ matrix with 
$ij^{\hbox{\eightpoint th}}$ entry the weight of the edge from vertex 
$i$ to vertex $j$:
$$
A_1 = \pmatrix{w_0 & w_1 \cr
               w_2 & w_3 \cr},
$$
where $w_{\alpha} := \llangle \alpha | \alpha \rrangle$ and $\alpha$ 
is the base 10 (say) representation of the local configuration bit
string.  The $ij^{\hbox{\eightpoint th}}$ entry in $A_1^n$ is the sum 
of the weights of the paths with $n$ edges from vertex $i$ to vertex 
$j$, so it is useful to define the generating function
$$
A_1(t) := \sum_{n \ge 0} A_1^n t^n.
$$
The $i^{\hbox{\eightpoint th}}$ diagonal entry in $A_1(t)$ is the sum 
of the weights of the paths of length $n$ beginning and ending at 
vertex $i$, times $t^n$, for all $n \ge 0$; hence $\Tr A_1(t)$ is the 
corresponding sum over all closed paths in $G_1$.  It is 
straightforward to show [28] that if $Z(t) := \det(I - t A)$ then
$$
\Tr A(t) = -{t Z'(t) \over Z(t)}.                           \eqno(4.1)
$$
Evaluating the righthand side of (4.1) for $A_1$ we find
$$
\Tr A_1(t) 
 = {(w_0 + w_3 + 2 w_1 w_2 t) t - 2 w_0 w_3 t^2
    \over
    1 - (w_0 + w_3 + w_1 w_2 t) t + w_0 w_3 t^2
   }.                                                       \eqno(4.2)
$$
We can read off the weights of the cycles in $G_1$ directly from the 
positive terms in the numerator and the negative terms in the 
denominator of (4.2):  $w_0$, $w_3$, and $w_1 w_2$.  (That these are 
the right terms becomes clear upon expanding (4.2) in powers of $t$:
$$
\Tr A_1(t)
 = (w_0 + w_3) t + 
   (2 w_1 w_2 - 2 w_0 w_3 + 
    w_0^2 + 2 w_0 w_3 + w_3^2) t^2 + O(t^3),
$$
where the cancelling coefficients of $t^2$ have been included to 
indicate the function of the other terms in (4.2).)  This is more
machinery than we need to find the cycles in $G_1$, of course, but it
will become useful in more complicated situations.
}

%--------|---------|---------|---------|---------|---------|---------|
For either a periodic or an infinite QCA, each of these cycles must 
have weight 1---this is condition ({\it i\/}) in the Unitarity 
Theorems.  Thus
$$
\eqalignno{
1 &= w_0                                                    &(4.3a)\cr
1 &= w_3                                                    &(4.3b)\cr
1 &= w_1 w_2.                                               &(4.3c)\cr
}
$$
That is, both $|00\rrangle$ and $|11\rrangle$ must have norm 1, while
$\llangle 10|10 \rrangle = \llangle 01|01 \rrangle^{-1}$.

%--------|---------|---------|---------|---------|---------|---------|
\topinsert
{}\vskip -\baselineskip
\vskip -\baselineskip
\vskip -\baselineskip
$$
\epsfxsize=\usewidth\epsfbox{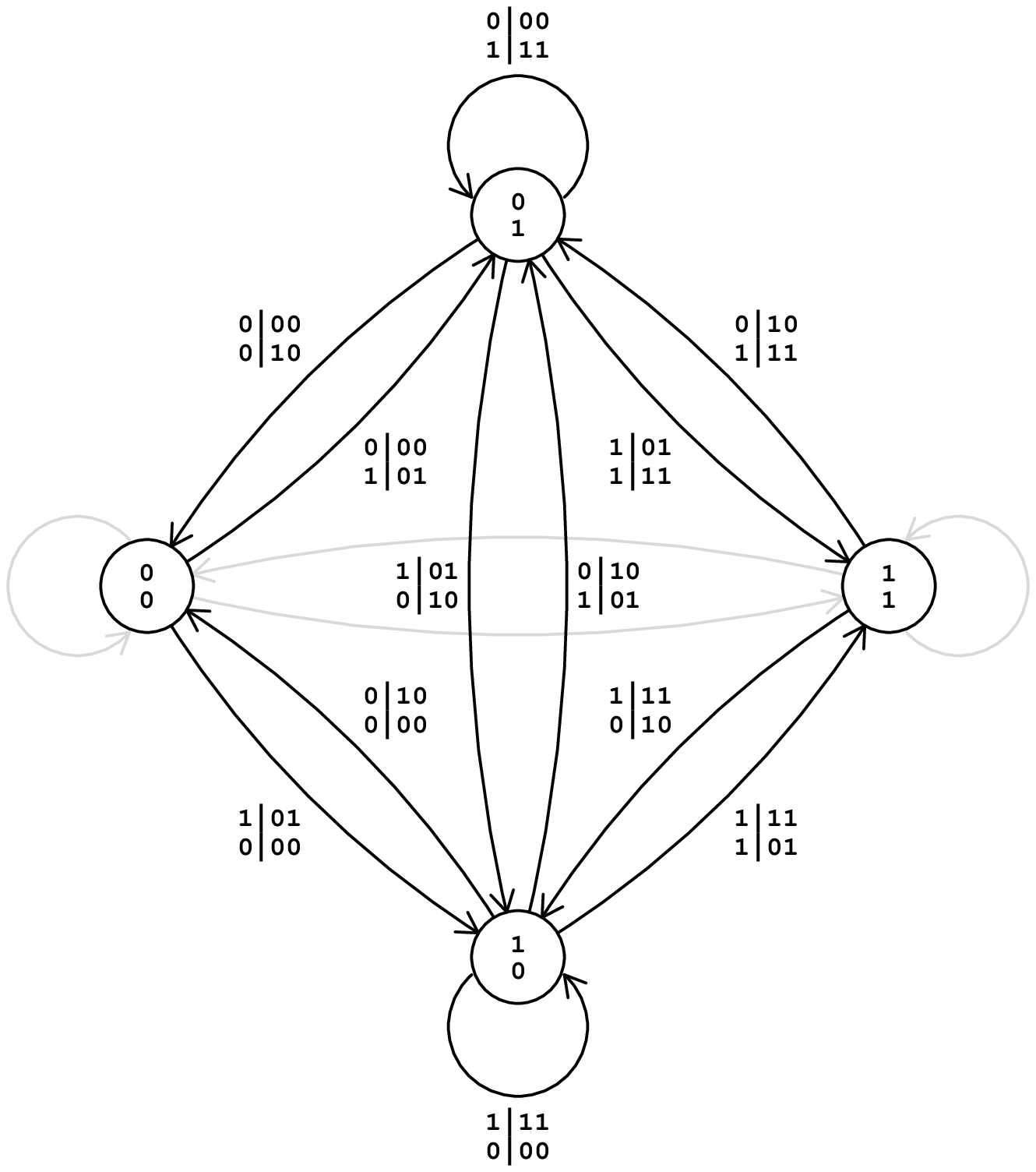}
$$
\vskip -\baselineskip
\vskip -\baselineskip
\vskip -\baselineskip
\eightpoint{%
{\narrower\noindent{\bf Figure 2}.  $G_2$ for binary CAs with local 
neighborhoods of size 2.  The edges in the subgraph isomorphic to 
$G_1$ are grey and unlabelled; each other edge is labelled by the pair
of numbers to the left of the slash and weighted by the inner product 
of the pair of amplitude vectors of the two local configurations to 
the right of the slash.\par}
}
\endinsert

%--------|---------|---------|---------|---------|---------|---------|
Figure 2 shows $G_2$ with only the edges in $M$ labelled and weighted; 
the subgraph isomorphic to $G_1$ is indicated with grey unlabelled 
edges.  Since the orthogonality constraints are determined by the
weights of paths in $M$, to simplify the transfer matrix we may set 
the weights of the edges from vertex $(0,0)$ to itself and from vertex
$(1,1)$ to itself to 0 and the weights of the edges between these two
vertices to 1.  Then
$$
A_2 = \pmatrix{   0    & w_{01} & w_{01} &   1    \cr
                w_{02} & w_{03} & w_{12} & w_{13} \cr
                w_{02} & w_{12} & w_{03} & w_{13} \cr
                  1    & w_{23} & w_{23} &   0    \cr},     \eqno(4.4)
$$
where $w_{\alpha\beta} := \llangle \alpha|\beta \rrangle$ and 
$\alpha,\beta$ are again base 10 (say) representations of the local 
configuration bit strings.  Now,
$$
\eqalign{
Z_2(t) 
 &:= \det(I - t A_2)                                               \cr
 &= 1 - 2w_{03}t 
    + (-1 - 2w_{01}w_{02} + w_{03}^2 - w_{12}^2 - 2w_{13}w_{23})t^2\cr
 &\quad + 2(w_{03} + w_{01}w_{02}w_{03} - w_{01}w_{02}w_{12} 
            - w_{01}w_{13}                                         \cr
 &\quad{\phantom{ + 2(w_{03}}} - w_{02}w_{23} + w_{03}w_{13}w_{23}
             - w_{12}w_{13}w_{23})t^3                              \cr
 &\quad + (-w_{03}^2 + w_{12}^2 + 2w_{01}w_{03}w_{13} 
           - 2w_{01}w_{12}w_{13} + 2w_{02}w_{03}w_{23} 
           - 2w_{02}w_{12}w_{23})t^4,                              \cr
}
$$
which we may use in (4.1) to conclude that the weights of the parts of
the cycles in $G_2$ which lie in $M$ are $w_{03}$, $w_{12}^2$, 
$w_{01}w_{02}$, $w_{13}w_{23}$, $w_{01}w_{13}$ and $w_{02}w_{23}$.

%--------|---------|---------|---------|---------|---------|---------|
The first two of these are weights of cycles entirely in $M$, while 
the last four are weights of acyclic paths in $M$ terminating at 
vertices of $G_1$.  Condition ({\it iii\/}) in the Unitarity Theorems
forces
$$
\eqalign{
0 &= w_{01}w_{02}                                                  \cr
0 &= w_{13}w_{23}                                                  \cr
0 &= w_{01}w_{13}                                                  \cr
0 &= w_{02}w_{23}.                                                 \cr
}                                                           \eqno(4.5)
$$
Recall that $w_{\alpha\beta} := \llangle \alpha|\beta \rrangle$, so 
that $w_{\alpha\beta} = 0$ means that $|\alpha\rrangle$ and 
$|\beta\rrangle$ are orthogonal.  Thus the only solutions to (4.5) 
satisfy exactly one of the following sets of relations:
$$
\eqalignno{
|00\rrangle \perp |01\rrangle
\; &\wedge \;
|10\rrangle \perp |11\rrangle                               &(4.6a)\cr
|00\rrangle \perp |10\rrangle
\; &\wedge \;
|01\rrangle \perp |11\rrangle,                              &(4.6b)\cr
}
$$
where $\wedge$ is the boolean relation `and'.  These two sets of 
relations transform into one another under the action of $P$; hence we 
need only consider one, say $(4.6a)$.

%--------|---------|---------|---------|---------|---------|---------|
To obtain a periodic QCA, condition ({\it ii\/}) of Theorem 3.9 must
also be satisfied, \ie, for the cycles in $M$:
$$
\eqalignno{
0 &= w_{03}                                                 &(4.7a)\cr
0 &= w_{12}^2.                                              &(4.7b)\cr
}
$$
Since the amplitude vectors for a binary CA lie in $\C^2$ and by (4.3) 
are nonzero, no more than two can be mutually orthogonal.  This means 
that the constraints (4.7) restrict $(4.6a)$ to the single frame:
$$
|00\rrangle \parallel |10\rrangle \>
 \perp \> |01\rrangle \parallel |11\rrangle,                \eqno(4.8)
$$
where by {\sl frame\/} we mean a pair $S_0$, $S_1$ of sets of vectors
in $\C^2$ such that each vector in $S_0$ is orthogonal to each vector 
in $S_1$.  We will denote the family of local rules satisfying the 
relations (4.8) subject to the normalization constraints (4.3) by
$f_{2,1}$:  the subscript 2 is $k$; the 1 indicates that local
configurations $i_{-k/2}\ldots i_{-1}i_1\ldots i_{k/2}$ with the same 
state $i_1$ have parallel amplitude vectors.  An explicit 
parameterization of this local rule gives the rule table for 
$f_{2,1}(i|\lambda)$:
$$
\eqalign{
\vbox{\offinterlineskip
\hrule height 0.8pt
\halign{%
\vrule#width 0.8pt&\strut\quad\hfil$#$\hfil\quad&%
\vrule#width 0.8pt&\strut\quad\hfil$#$\quad&%
\vrule#width 0.4pt&\strut\quad\hfil$#$\quad&%
\vrule#width 0.8pt\cr
height2pt&\omit&&\omit&&\omit&\cr
&f_{2,1}&&0\hfil&&1\hfil&\cr
height2pt&\omit&&\omit&&\omit&\cr
\noalign{\hrule height 0.8pt}
height2pt&\omit&&\omit&&\omit&\cr
&00&&e^{i\alpha}\cos\theta&&%
     e^{i\beta}i\sin\theta&\cr
height2pt&\omit&&\omit&&\omit&\cr
\noalign{\hrule}
height2pt&\omit&&\omit&&\omit&\cr
&01&&e^{i\phi_1}\rho e^{-i\beta}i\sin\theta&&%
     e^{i\phi_1}\rho e^{-i\alpha}\cos\theta&\cr
height2pt&\omit&&\omit&&\omit&\cr
\noalign{\hrule}
height2pt&\omit&&\omit&&\omit&\cr
&10&&e^{i\phi_2}\rho^{-1} e^{i\alpha}\cos\theta&&%
     e^{i\phi_2}\rho^{-1} e^{i\beta}i\sin\theta&\cr
height2pt&\omit&&\omit&&\omit&\cr
\noalign{\hrule}
height2pt&\omit&&\omit&&\omit&\cr
&11&&e^{-i\beta}i\sin\theta&&%
     e^{-i\alpha}\cos\theta&\cr
height2pt&\omit&&\omit&&\omit&\cr}
\hrule height 0.8pt}
&\cr}                                                       \eqno(4.9)
$$
where $\lambda$ labels the rows and $i$ the columns.  The four entries
in rows 00 and 11 form an arbitrary $SU(2)$ matrix: 
$\alpha, \beta, \theta \in [0,2\pi)$; the remaining degree of freedom
is an overall phase which has been divided out as it has no effect on
probabilities.  $|10\rrangle$ is parallel to $|00\rrangle$, differing 
by an arbitrary factor $\rho^{-1} e^{i\phi_2} \in \C$.  $|01\rrangle$ 
is parallel to $|11\rrangle$ with length $\rho$ and phase angle 
$\phi_1 \in [0,2\pi)$.  The only other possible periodic local rule, 
satisfying $(4.6b)$, is the the parity transform:  
$f_{2,-1} = Pf_{2,1}$, obtained by interchanging the middle two rows 
in (4.9).  The state transposition $\tau$ leads to no new rules:  
$\tau_{\rm in}$ is implemented by 
$\alpha \to \pi/2 - \beta$, $\beta \to \pi/2 - \alpha$, 
$\theta \to \pi/2 - \theta$, $\phi_1 \leftrightarrow \phi_2$ and 
$\rho \to \rho^{-1}$, while $\tau_{\rm out}$ is implemented just by
$\alpha \to \pi/2 + \beta$, $\beta \to \pi/2 + \alpha$ and
$\theta \to \pi/2 - \theta$. 

%--------|---------|---------|---------|---------|---------|---------|
To obtain an infinite QCA, by Corollary 3.5 and Theorem 3.6, a local 
rule $f$ must have a nonempty deterministic sector consisting of local 
configurations corresponding to the edges in a collection of cycles in
$G_1$.  The simplest possibility is that 0 is a quiescent state, so 
that $|00\rrangle = (1,0)$.  Conditions ({\it ii\/}) and ({\it iv\/}) 
of Theorem 3.10 impose no additional constraints beyond (4.3) and 
(4.5) if 00 is the only local configuration in $D_{\! f}$; using 
$(4.6a)$ again we find the local rule $f_{2,1;00}$ (where the 
subscript 00 is the deterministic sector), which may be parameterized 
as:
$$
\eqalign{
\vbox{\offinterlineskip
\hrule height 0.8pt
\halign{%
\vrule#width 0.8pt&\strut\quad\hfil$#$\hfil\quad&%
\vrule#width 0.8pt&\strut\quad\hfil$#$\quad&%
\vrule#width 0.4pt&\strut\quad\hfil$#$\quad&%
\vrule#width 0.8pt\cr
height2pt&\omit&&\omit&&\omit&\cr
&f_{2,1;00}&&0\hfil&&1\hfil&\cr
height2pt&\omit&&\omit&&\omit&\cr
\noalign{\hrule height 0.8pt}
height2pt&\omit&&\omit&&\omit&\cr
&00&&1&&%
     0&\cr
height2pt&\omit&&\omit&&\omit&\cr
\noalign{\hrule}
height2pt&\omit&&\omit&&\omit&\cr
&01&&0&&%
     e^{i\phi_1}\rho&\cr
height2pt&\omit&&\omit&&\omit&\cr
\noalign{\hrule}
height2pt&\omit&&\omit&&\omit&\cr
&10&&\rho^{-1} e^{i\alpha}\cos\theta&&%
     \rho^{-1} e^{i\beta}i\sin\theta&\cr
height2pt&\omit&&\omit&&\omit&\cr
\noalign{\hrule}
height2pt&\omit&&\omit&&\omit&\cr
&11&&e^{i\phi_3}e^{-i\beta}i\sin\theta&&%
     e^{i\phi_3}e^{-i\alpha}\cos\theta&\cr
height2pt&\omit&&\omit&&\omit&\cr}
\hrule height 0.8pt}
&\cr}                                                       \eqno(4.10)
$$
since $|01\rrangle$ is orthogonal to $|00\rrangle$, with arbitrary
nonzero component $\rho e^{i\phi_1} \in \C$, while the entries in
rows 10 and 11 would form an arbitrary $U(2)$ matrix $\Phi^{(1)}$ but 
for the length $\rho^{-1}$ of $|10\rrangle$:  
$\alpha, \beta, \phi_1, \phi_3, \theta \in [0,2\pi)$ and the overall
phase has been set by the choice $|00\rrangle = (1,0)$.  Finally, 
surjectivity constraint ({\it i\/}) of Theorem 3.18 requires in 
addition that $f(0|10) \not= 0$, \ie, $\cos\theta \not= 0$; 
surjectivity constraint ({\it ii\/}) is already satisfied because of 
the orthogonality relations.  Again, the parity transform 
$f_{2,-1;00} = Pf_{2,1;00}$ is also a solution.

%--------|---------|---------|---------|---------|---------|---------|
Condition ({\it ii\/}) of Theorem 3.10 will impose additional 
constraints only if the vertex labelled 1 in Figure 1 is also in 
$D_1$.  This occurs if $\{00,11\} \subset D_{\! f}$ or 
$\{00,01,10\} \subset D_{\! f}$.  In either case $w_1 = 1 = w_2$, 
since $\ldots 01 \ldots$ and $\ldots 10 \ldots$ are admissible 
configurations in the former and $\ldots 0101 \ldots$ is an admissible
configuration in the latter.  Furthermore, in the first case 
$D_2 \cap M$ contains the cycles with weights $w_{03}$ and in the 
second case the one with weight $w_{01}^2$, so condition ({\it iv\/}) 
of Theorem 3.10 imposes constraint $(4.7a)$ or $(4.7b)$, respectively,
either of which restricts $(4.6a)$ to (4.8).  This leaves only the 
completely deterministic limit 
$\theta = 0 = \alpha = \phi_1 = \phi_2$, $\rho = 1$ of the periodic 
rule (4.9), which we will denote by $f_{2,1;D}$ (the subscript $D$ 
indicates that the rule is completely deterministic), together with 
its parity transform.  Injectivity implies surjectivity in this case
[32,33], so Theorem 3.18 imposes no additional constraints.

%--------|---------|---------|---------|---------|---------|---------|
Starting with 0 being an `anti-quiescent' state, \ie, 
$|00\rrangle = (0,1)$, forces $11 \in D_{\! f}$ immediately and leads 
by exactly the same argument to the limit $\theta = \pi/2 = \beta, 
\phi_1 = \phi_2 = 0, \rho = 1$ of (4.9), which is again completely 
deterministic and is immediately identifiable as both 
$\tau_{\rm in} f_{2,1;D}$ and $\tau_{\rm out} f_{2,1;D}$.

%--------|---------|---------|---------|---------|---------|---------|
This discussion could be repeated starting with 1 as a `quiescent' 
state and would lead to the $\tau_{\rm in}\tau_{\rm out}$ 
transformation of the infinite QCA rules found in the previous three
paragraphs.

%--------|---------|---------|---------|---------|---------|---------|
The last possibility for an infinite QCA is for $D_{\! f}$ to contain
only the edges 01 and 10 which form a cycle in $G_1$.  By condition 
({\it iv\/}) of Theorem 3.10, $(4.7b)$ must be satisfied, \ie,
$\llangle 01|10 \rrangle = 0$, so again $(4.6a)$ is restricted to 
(4.8) and we are left with exactly the same completely deterministic 
limits already discussed.  

%--------|---------|---------|---------|---------|---------|---------|
Note that these four deterministic CAs:  $f_{2,1;D}$, $Pf_{2,1;D}$, 
$\tau_{\rm in} f_{2,1;D}$ and $P\tau_{\rm in} f_{2,1;D}$, are trivial 
in the sense that the local rule depends on only one of the cell 
states in the local configuration, \ie, 
$f_{2,1;D}(i|i_1i_2) = \delta_{ii_2}$.  That the only deterministic 
binary CAs with neighborhoods of size 2 or 3 are trivial in this sense 
is well known [35] (this is also the sense in which the only linear 
QCAs which are allowed by the No-Go Theorem are trivial for 
{\sl any\/} size neighborhood [8]); we have just shown that if 
$D_{\! f}$ contains more than just 00 (or 11) then infinite $k = 2$ 
QCAs are trivial in the same way.  It is most interesting that when
$D_{\! f} = \{00\}$ (or $\{11\}$) the local rule is not completely
deterministic and, furthermore, partitions the amplitude vectors into
two independent frames rather than the single frame of the periodic
local rules.

%--------|---------|---------|---------|---------|---------|---------|
\topinsert
{}\vskip -\baselineskip
\vskip -\baselineskip
\vskip -\baselineskip
$$
\epsfxsize=\usewidth\epsfbox{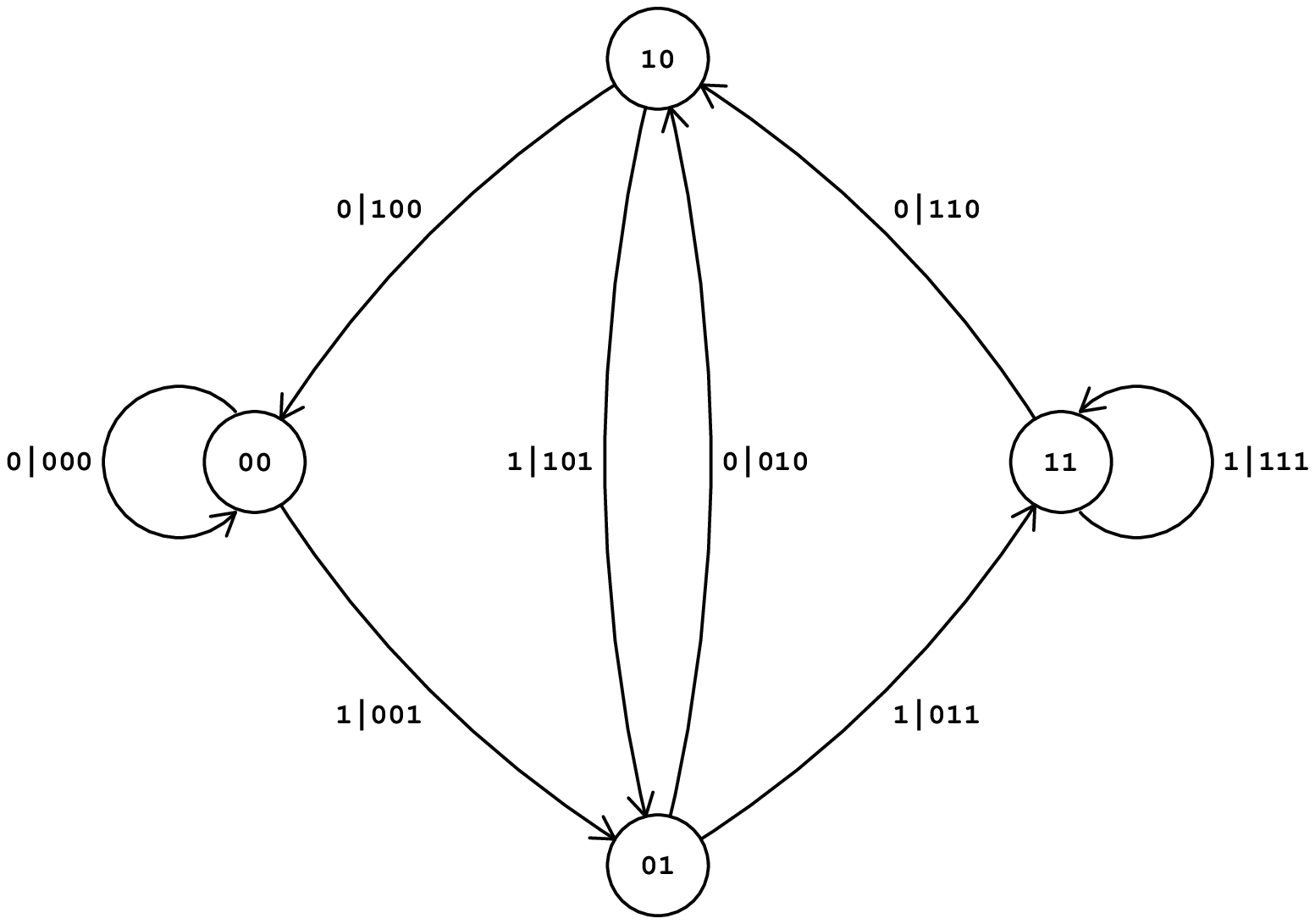}
$$
\vskip -\baselineskip
\vskip -\baselineskip
\eightpoint{%
\centerline{{\bf Figure 3}.  $G_1$ for binary CAs with local 
neighborhoods of size 3.}
}
\endinsert

\noindent{\it 4.2.  $k = 3$}
\nobreak

\nobreak
%--------|---------|---------|---------|---------|---------|---------|
The next simplest binary CAs have local neighborhoods of size 3.
Figure 3 shows $G_1$ (in this subsection the arguments $Q = \{0,1\}$
and $k = 3$ are suppressed).  The cycles may be determined as in the
previous subsection; then condition ({\it i\/}) of the Unitarity 
Theorems imposes the constraints:
$$
\eqalign{1 &= w_0  \cr
         1 &= w_7  \cr}
\qquad
\eqalign{1 &= w_2w_5     \cr
         1 &= w_1w_2w_4  \cr}
\qquad
\eqalign{1 &= w_3w_6w_5     \cr
         1 &= w_1w_3w_6w_4. \cr}                           \eqno(4.11)
$$
Note that only the first five of these constraints are independent:
reading the columns left to right, the last equation is implied by the 
three preceding.

%--------|---------|---------|---------|---------|---------|---------|
$G_2$ is quite complicated to draw, but by computing successive powers 
of $A_2$ (now a $16 \times 16$ matrix, so computing the determinant in 
$Z_2(t)$ is tedious, even by computer), we find the weights of the 
parts of the acyclic paths of length $n \le 4$ in $M$ terminating at 
vertices of $G_1$; they are listed in Appendix A.  Condition 
({\it iii\/}) in the Unitarity Theorems requires that each of these 
weights vanish.  The only solutions to these constraints satisfy one 
of the following sets of relations:
$$
\eqalignno{
|000\rrangle \perp |001\rrangle
 \; \wedge \; 
 |010\rrangle \perp |011\rrangle
 \; &\wedge \;
 |100\rrangle \perp |101\rrangle
 \; \wedge \; 
 |110\rrangle \perp |111\rrangle                           &(4.12a)\cr
|000\rrangle \perp |100\rrangle
 \; \wedge \; 
 |010\rrangle \perp |110\rrangle
 \; &\wedge \;
 |001\rrangle \perp |101\rrangle
 \; \wedge \; 
 |011\rrangle \perp |111\rrangle                           &(4.12b)\cr
|000\rrangle \parallel |100\rrangle \perp
 |010\rrangle \parallel |110\rrangle
 \; &\wedge \;
 |001\rrangle \parallel |101\rrangle \perp
 |011\rrangle \parallel |111\rrangle                       &(4.12c)\cr
|000\rrangle \parallel |001\rrangle \perp
 |010\rrangle \parallel |011\rrangle
 \; &\wedge \;
 |100\rrangle \parallel |101\rrangle \perp
 |110\rrangle \parallel |111\rrangle.                      &(4.12d)\cr
}
$$
That any of these sets of relations implies that the weights of 
{\sl all\/} acyclic paths of {\sl any\/} length in $M$ terminating at 
vertices of $G_1$ vanish is a consequence of Theorem 5.1 (which is 
stated and proved in the next section).  Just as in the $k = 2$ case, 
the second and fourth sets of relations are merely the parity 
transforms of the first and third, respectively, so we need only 
consider the possibilities $(4.12a)$ and $(4.12c)$.

%--------|---------|---------|---------|---------|---------|---------|
To obtain a periodic QCA, condition ({\it ii\/}) of Theorem 3.9 must
also be satisfied, \ie, we must consider the cycles in $M$.  There are
four cycles of length $n \le 2$, with weights:  $w_{07}$, 
$w^2_{25}$, $w_{02}w_{05}$ and $w_{27}w_{57}$.  Of the eight cycles of
length $n = 3$, four are contained entirely in $M$, with weights:  
$w_{03}w_{05}w_{06}$, $w_{12}w_{14}w_{24}$, $w_{17}w_{27}w_{47}$ 
and $w_{35}w_{36}w_{56}$; the other four are acyclic paths terminating 
at vertices in $G_1$ and so are already included in the list in
Appendix A.  Imposing the constraint that each of these weights vanish 
implies that the relations $(4.12a)$ are restricted to a single frame:
$$
\eqalignno{
|000\rrangle \parallel |010\rrangle \parallel
|100\rrangle \parallel |110\rrangle &\perp
|001\rrangle \parallel |011\rrangle \parallel
|101\rrangle \parallel |111\rrangle,                       &(4.13a)\cr
\noalign{\smallskip\smallskip
\hbox{while the relations $(4.12c)$ are restricted to the frame:}
\smallskip\smallskip}
|000\rrangle \parallel |001\rrangle \parallel
|100\rrangle \parallel |101\rrangle &\perp
|010\rrangle \parallel |011\rrangle \parallel
|110\rrangle \parallel |111\rrangle.                       &(4.13c)\cr
}
$$
Note that $(4.13c)$ is invariant under parity reversal (and hence is 
also the consequence of $(4.12d)$), while $(4.13a)$ is not (and hence
$(4.12b)$ leads to its parity transform).  These three frames are the 
only possibilities for local rules.  To show that the longer cycles in
$G_2$ rule out none of them, consider one and write the transition 
matrix $A_2$, as in (4.4), with 0s for the weights which vanish in the 
given frame.  The resulting matrix is sufficiently sparse that 
$Z_2(t) = \det(I - t A_2)$ can be computed easily.  In each of these 
frames we find that $Z_2(t) = 1 - t^2 - 2t^3 - t^4$.  Since we may 
compute $\Tr A_2(t)$ from $Z_2(t)$ by (4.1), this means that there are 
no further orthogonality constraints on the weights.

%--------|---------|---------|---------|---------|---------|---------|
Thus $(4.13a)$ subject to the normalization constraints (4.11) gives a 
local rule $f_{3,1}$ (for $k$ odd we label the local configuration
$i_{-(k-1)/2}\ldots i_0\ldots i_{(k-1)/2}$; the subscript 1 again
indicates that the frame in (4.13a) partitions the amplitude vectors 
according to $i_1$ in the local configuration).  A more concise 
description of this local rule than an explicit parameterization like
(4.9) is that the amplitude vectors for $f_{3,1}$ satisfy:
\global\setbox4=\hbox to 268.96184pt{%
 $z_i \in \C, i \in \{1,\ldots,6\}$; 
 $|z_2z_5| = 1$, $|z_1z_2z_4| = 1$, $|z_3z_5z_7| = 1$.\hfil}
$$
\eqalign{
&\hbox{$|000\rrangle$ and $|111\rrangle$ 
       form an orthonormal basis of $\C^2$}                        \cr
&\hbox{$|010\rrangle = z_2|000\rrangle$,
       $|100\rrangle = z_4|000\rrangle$,
       $|110\rrangle = z_6|000\rrangle$}                           \cr
&\hbox{$|001\rrangle = z_1|111\rrangle$,
       $|011\rrangle = z_3|111\rrangle$,
       $|101\rrangle = z_5|111\rrangle$}                           \cr
&\box4                                                             \cr
}                                                         \eqno(4.14a)
$$
The amplitude vectors satisfying the constraints $(4.13c)$ subject to 
the normalization constraints (4.11) may be described similarly, 
changing the middle two lines of $(4.14a)$ to:
$$
\global\setbox5=\hbox to 268.96184pt{%
 $|001\rrangle = z_1|000\rrangle$,
 $|100\rrangle = z_4|000\rrangle$,
 $|101\rrangle = z_5|000\rrangle$\hfil}
\eqalign{
&\box5                                                             \cr
&\hbox{$|010\rrangle = z_2|111\rrangle$,
       $|011\rrangle = z_3|111\rrangle$,
       $|110\rrangle = z_6|111\rrangle$.}                          \cr
}                                                         \eqno(4.14c)
$$
We denote this rule by $f_{3,0}$.  Finally, the parity transform
$f_{3,-1} = Pf_{3,1}$ gives the solution corresponding to $(4.12b)$.
Each of these rules may be parameterized by 12 real parameters and an
overall phase:  6 for the first pair of orthogonal vectors, 2 for each
additional vector in the frame, and $-1$ for each independent 
normalization constraint.  These are all the rule families for the 
periodic case; just as when $k = 2$ the state transposition $\tau$ 
leads to no additional rules.

%--------|---------|---------|---------|---------|---------|---------|
For an infinite QCA, the possible deterministic sectors are determined
by the cycles in $G_1$.  Viewing Figure 3 or recalling (4.11), we see
that the sets of local configurations appearing in cycles are:
$\{000\}$, $\{111\}$, $\{010,101\}$, $\{001,010,001\}$, 
$\{011,101,110\}$ and $\{001,011,100,110\}$; any union of one or more 
of these sets is a possibility for $D_{\! f}$.

%--------|---------|---------|---------|---------|---------|---------|
The simplest possibility is $D_{\! f} = \{000\}$, \ie, 0 is uniquely
quiescent.  In this case $|000\rrangle = (1,0)$ and conditions 
({\it ii\/}) and ({\it iv\/}) of Theorem 3.10 impose no additional 
constraints beyond (4.11) and (4.12).  $(4.12a)$ defines the rule 
family denoted $f_{3,1;000}$ with amplitude vectors partitioned into
four frames:
$$
\global\setbox6=\hbox to 268.96184pt{%
 $|000\rrangle = (1,0)$; 
 $|001\rrangle = (0,z_1)$\hfil}
\eqalign{
&\box6                                                             \cr
&\hbox{$|\alpha0\rrangle$ and $|\alpha1\rrangle$ are orthogonal for
       $\alpha \in \{01,10,11\}$}                                  \cr
&\hbox{the norms of the amplitude vectors satisfy (4.11),}         \cr
}                                                         \eqno(4.15a)
$$
a 16 real parameter family of local rules.  Surjectivity constraint 
({\it i\/}) of Theorem 3.18 rules out only the codimension 1
submanifolds defined by $f(0|010) = 0$, $f(0|100) = 0$ and 
$f(0|110) = 0$, while surjectivity constraint ({\it ii\/}) is again 
satisfied as a consequence of the orthogonality relations.  Similarly, 
$(4.12c)$ would define a rule family with amplitude vectors 
satisfying:
$$
\global\setbox7=\hbox to 268.96184pt{%
 $|000\rrangle = (1,0)$; 
 $|001\rrangle = (z_1,0)$; 
 $|010\rrangle = (0,z_2)$; 
 $|011\rrangle = (0,z_3)$\hfil}
\eqalign{
&\box7                                                             \cr
&\hbox{$|10i\rrangle$ and $|11j\rrangle$ are orthogonal for 
       $i,j \in \{0,1\}$}                                          \cr
&\hbox{the norms of the amplitude vectors satisfy (4.11),}         \cr
}                                                         \eqno(4.15c)
$$
except that $|010\rrangle = (0,z_2)$ conflicts with the surjectivity
constraint that $f(0|010) = 0$, ruling out this possibility.  Thus 
$f_{3,1;000}$ and its distinct parity transform are the only allowed 
rule families when $D_{\! f} = \{000\}$.  Both of these infinite QCAs 
with 0 uniquely quiescent have distinct state transposition transforms 
under $\tau_{\rm in}\tau_{\rm out}$ with 1 as a `quiescent' state.

%--------|---------|---------|---------|---------|---------|---------|
The next simplest possibility is that $D_{\! f} = \{000,111\}$.  Then
by condition ({\it ii\/}) of Theorem 3.10, the two acylic paths in 
$G_1$ terminating at 000 and 111 must each have weight 1:
$$
\eqalign{
1 &= w_1w_3                                                        \cr
1 &= w_4w_6.                                                       \cr
}                                                          \eqno(4.16)
$$
With (4.11), the second of these constraints is implied by the first.
By condition ({\it iv\/}) of Theorem 3.10, the cycles in $D_2 \cap M$
must have vanishing weights.  Hence $0 = w_{07}$.  This constraint 
restricts $(4.12a)$ to the three frame set of relations:
$$
|000\rrangle \parallel |110\rrangle \perp
|001\rrangle \parallel |111\rrangle 
 \; \wedge \;
|010\rrangle \perp |011\rrangle 
 \; \wedge \;
|100\rrangle \perp |101\rrangle                           \eqno(4.17a)
$$
and restricts $(4.12c)$ to $(4.13c)$.  We define the rule 
$f_{3,1;000,111}$ by the set of amplitude vectors satisfying $(4.17a)$ 
subject to the normalization constraints (4.11) and (4.16) and with 
$D_{\! f} = \{000,111\}$, described by:
$$
\global\setbox8=\hbox to 268.96184pt{%
 $|000\rrangle = (1,0)$; 
 $|110\rrangle = (z_3,0)$,
 $|001\rrangle = (0,z_1)$; 
 $|111\rrangle = (0,1)$\hfil}
\eqalign{
&\box8                                                             \cr
&\hbox{$|\alpha0\rrangle$ and $|\alpha1\rrangle$ are orthogonal for
       $\alpha \in \{01,10\}$}                                     \cr
&\hbox{the norms of the amplitude vectors satisfy (4.11) and 
       (4.16),}                                                    \cr
}                                               \eqno\phantom{(4.18a)}
$$
a 12 real parameter family of local rules.  Surjectivity condition 
({\it ii\/}) of Theorem 3.18 restricts these parameter values by 
removing the codimension 1 submanifolds defined by $f(0|010) = 0$, 
$f(0|100) = 0$, $f(1|011) = 0$ and $f(1|101) = 0$.  Surjectivity 
condition ({\it ii\/}) is ensured by the orthogonality of 
$|\alpha 0\rrangle$ and $|\alpha 1\rrangle$ for all $\alpha \in Q^2$.  
$f_{3,1;000,111}$ has a distinct parity transform but is invariant 
under $\tau_{\rm in}\tau_{\rm out}$.  Applying either of the state 
transposition transforms alone produces a distinct rule family in
which 0 is `anti-quiescent', as is 1.

%--------|---------|---------|---------|---------|---------|---------|
The rule which would be defined by $(4.14c)$ with 
$|000\rrangle = (1,0)$ and $|111\rrangle = (0,1)$ is ruled out by 
surjectivity condition ({\it i\/}) of Theorem 3.18:  
$|010\rrangle = (0,z_2)$ contradicts $f(0|010) \not= 0$, for example.
Similarly, $|000\rrangle = (0,1)$ and $|111\rrangle = (1,0)$ fails to
be surjective:  in this case $|010\rrangle = (z_2,0)$ contradicts 
$f(1|010) \not= 0$.

%--------|---------|---------|---------|---------|---------|---------|
We may continue to increase the size of $D_{\! f}$ and find further
infinite QCA rules; the procedure is clear.  When the deterministic 
sector gets only a little larger, there will be only completely 
deterministic rules, just as in the $k = 2$ case of the previous 
subsection.

\medskip
\noindent{\bf 5.  Discussion}

%--------|---------|---------|---------|---------|---------|---------|
\noindent The results of the previous section demonstrate that the 
Unitarity Theorems provide an effective procedure for finding one
dimensional binary QCAs, both periodic and infinite.  Although it is
increasingly difficult to find the most general unitary solutions for 
large local neighborhood size $k$---the procedure is not very
efficient---the results for $k = 2$ and $k = 3$ suggest a pattern for
some {\sl particular\/} solutions.  Specifically, the sets of 
relations (4.6) and (4.12) generalize to larger values of $k$.  If we
generalize the notion of `frame' to $q$-{\sl frame\/}:  a collection
$S_0,\ldots,S_{q-1}$ of sets of vectors in $\C^q$ such that each 
vector in $S_i$ is orthogonal to each vector in $S_j$ for $i \not= j$,
we can state the following:

%--------|---------|---------|---------|---------|---------|---------|
\noindent\Theorem\ 5.1.  {\sl Let $0 < j < k \in \Z$.  For each 
$\gamma \in Q^j$ define a $q$-frame
$$
\eqalignno{
S^{(\gamma)} &= \{S_0^{(\gamma)},\ldots,S_{q-1}^{(\gamma)}\}       \cr
\noalign{\hbox{by}}
S_i^{(\gamma)} &= 
 \{|i_i\ldots i_k\rrangle \mid 
   i_1\ldots i_j = \gamma, i_{j+1} = i \in Q\}.                    \cr
}
$$
Then condition ({\it iii\/}) of the Unitarity Theorems---the weight of
any acyclic path in $M(Q,k)$ terminating at vertices in $\Gone$ 
vanishes---holds when the amplitude vectors are partitioned into the 
$q^j$ frames $\{S^{(\gamma)}\}$, or into their parity transforms.}

%--------|---------|---------|---------|---------|---------|---------|
\noindent\Proof.  Consider any acyclic path in $M(Q,k)$ starting from
a vertex $(\alpha,\alpha) \in \Gone \subset \Gtwo$, where 
$\alpha \in Q^{k-1}$.  The $(k-j)^{\hbox{\eightpoint th}}$ edge in the
path is necessarily labelled 
$$
(i^{\vphantom{'}}_1\ldots i^{\vphantom{'}}_ji''_{j+1}\ldots i''_k,
 i^{\vphantom{'}}_1\ldots i^{\vphantom{'}}_ji'_{j+1} \ldots i'_k),
$$
where $i_1\ldots i_j$ are the rightmost $j$ states in $\alpha$ and
$i''_{j+1} \not= i'_{j+1}$ since the first edge of the path would not 
lie in $M(Q,k)$ otherwise.  Let $\gamma = i_1\ldots i_j$; then
$|i^{\vphantom{'}}_1\ldots i^{\vphantom{'}}_j
  i''_{j+1}\ldots i''_k\rrangle \in S_{i''_{j+1}}^{(\gamma)}$ and
$|i^{\vphantom{'}}_1\ldots i^{\vphantom{'}}_j
  i'_{j+1} \ldots i'_k\rrangle \in S_{i'_{j+1}}^{(\gamma)}$.  Since 
all the amplitude vectors in $S_{i''}^{(\gamma)}$ are orthogonal to 
those in $S_{i'}^{(\gamma)}$ for $i'' \not= i'$, 
$$
\llangle i^{\vphantom{'}}_1\ldots i^{\vphantom{'}}_j
         i''_{j+1}\ldots i''_k|
 i^{\vphantom{'}}_1\ldots i^{\vphantom{'}}_j
 i'_{j+1} \ldots i'_k\rrangle = 0
$$
and the weight of the path vanishes.  To show that the parity 
transformed set of frames $P\{S^{(\gamma)}\}$ also enforces condition
({\it iii\/}) of the Unitarity Theorems, make the analogous argument
using the terminal vertex of the acyclic path rather than its initial
vertex.                                                \hfill\endproof

%--------|---------|---------|---------|---------|---------|---------|
The special case $q = 2$ and $k = 3$ of Theorem 5.1 shows that each of
the sets of relations (4.12) which were found by considering only 
acyclic paths of length $n \le 4$ implies that the weights of all 
acyclic paths of any length in $M(\{0,1\},3)$ terminating at vertices 
of $G_1(\{0,1\},3)$ vanish; $(4.12a)$ and $(4.12b)$ are parity dual
sets of frames for $j = 2$, while $(4.12c)$ and $(4.12d)$ are the 
parity dual sets of frames for $j = 1$.

%--------|---------|---------|---------|---------|---------|---------|
That the converse of Theorem 5.1 is false is demonstrated by the 
existence of the nontrivial $k = 4$ {\sl reversible\/} deterministic
CA found by Patt [36], with local rule:
$$
f(i|i_1i_2i_3i_4) = 
 \cases{1 - \delta_{ii_2} & if $i_1 = i_4 = 0$, $i_3 = 1$;         \cr
        \delta_{ii_2}     & otherwise.                             \cr
       }
$$
Reversible CAs are {\it a fortiori\/} unitary and this local rule 
partitions the $k = 4$ amplitude vectors inconsistently with each of
the sets of frames described in Theorem 5.1.

%--------|---------|---------|---------|---------|---------|---------|
Consideration of this example leads to the observation that any 
reversible deterministic CA can be `quantized':  The local rule of 
such a CA partitions the amplitude vectors $|i_1\ldots i_k\rrangle$ 
into a single $q$-frame according to the unique $i \in Q$ for which
$f(i|i_1\ldots i_k)$ is nonzero.  Any rigid rotation of $\C^q$ 
preserves this $q$-frame, and hence unitarity, but gives, generically, 
nonzero transition amplitudes for all the $f(i|i_1\ldots i_k)$.  The
resulting global evolution is unitarily inequivalent to the original
reversible deterministic evolution.

%--------|---------|---------|---------|---------|---------|---------|
Although the local rules for the periodic QCAs found in Section 4 also
partition the amplitude vectors into a single frame (see (4.8) and 
(4.13)), they have additional degrees of freedom associated with the 
lengths of the amplitude vectors:  1 when $k = 2$ and 3 when $k = 3$; 
this should be contrasted with deterministic local rules for which all 
the amplitude vectors have length 1.  Despite being asymptotically
deterministic, the infinite QCAs with local rules found in Section 4 
are even further from the deterministic situation; their amplitude 
vectors lie in more than a single frame:  as many as $2^{k-1}$ for 
some of the QCAs with 0 uniquely quiescent (see (4.10) and $(4.15a)$).

%--------|---------|---------|---------|---------|---------|---------|
The multidimensionality of the local rule spaces for even the small 
neighborhood QCAs we have considered suggests that binary QCAs may 
have a wide range of quantum behaviors/computational power.  Whether 
any are computationally universal remains to be discovered.  There is
a long standing conjecture that computational power will be maximal at 
critical points of a physical theory [37].  Since the rule spaces here 
are smoothly parameterized this is a more natural arena in which to 
investigate this conjecture than is the deterministic case.  

%--------|---------|---------|---------|---------|---------|---------|
Consideration of QCAs as physical models, possibly with critical
points, raises the question of the continuum limits of these models.
The simplest nontrivial%
\sfootnote*{Recall that the No-go Theorem requires that the QCA not be
homogeneous if it is to be nontrivial.}
one dimensional linear binary QCAs have the $1+1$ dimensional Dirac 
equation as their continuum limit [9].  From the perspective of 
fundamental physics, it would be most interesting to determine the
continuum limits of the simple nonlinear models we have found here and
to extend them to higher dimensions.  That the reversible 
deterministic billiard ball model is computational universal [38]
suggests that higher dimensional QCAs might also be easier to prove 
computationally powerful.  It should be noted, however, that there can
be no analogue of the Unitarity Theorems in higher dimensions since
reversibility of deterministic CAs is undecidable in two dimensions 
[39]; the best we can expect is, as in Theorem 5.1, to find particular 
sets of local rules which ensure unitarity.

%--------|---------|---------|---------|---------|---------|---------|
Despite the existence of computation universal deterministic CAs,
probably their most important applications are simulations of physical
systems [40].  Similarly, it seems likely that QCAs will prove 
optimally suited not to universal computation but for the simulation 
of specific quantum mechanical systems and the solution of particular 
classes of problems.
\medskip

\noindent{\bf Acknowledgements}

%--------|---------|---------|---------|---------|---------|---------|
\noindent It is a pleasure to thank Peter Doyle and Michael Freedman
for asking the questions which led me to think about nonlinear QCAs;
Francis Zane for showing me Watrous' paper [12]; Peter Monta for 
telling me about convolutional codes (see, \eg, [30]); Ian Agol, Scott 
Crass and Wendy Miller for a conversation about generating functions; 
Brosl Hasslacher for discussions about computational power of QCAs; 
and Christoph D\"urr for sending me a preprint of [18].
\medskip

\vfill\eject
\medskip
\noindent{\bf Appendix A}

%--------|---------|---------|---------|---------|---------|---------|
\noindent The weights of the acyclic paths of length $n \le 4$ in
$M(\{0,1\},3)$ terminating at vertices of $G_1(\{0,1\},3)$ are:

\noindent $n = 3$:
$$
\eqalign{
&w_{01}w_{02}w_{04}\phantom{w_{00}}\quad
 w_{01}w_{02}w_{15}\phantom{w_{00}}\quad
 w_{01}w_{13}w_{26}\phantom{w_{00}}\quad
 w_{01}w_{13}w_{37}\phantom{w_{00}}                                \cr
&w_{02}w_{04}w_{45}\phantom{w_{00}}\quad
 w_{02}w_{15}w_{45}\phantom{w_{00}}\quad
 w_{13}w_{26}w_{45}\phantom{w_{00}}\quad
 w_{13}w_{37}w_{45}\phantom{w_{00}}                                \cr
&w_{04}w_{23}w_{46}\phantom{w_{00}}\quad
 w_{15}w_{23}w_{46}\phantom{w_{00}}\quad
 w_{23}w_{26}w_{57}\phantom{w_{00}}\quad
 w_{23}w_{37}w_{57}\phantom{w_{00}}                                \cr
&w_{04}w_{46}w_{67}\phantom{w_{00}}\quad
 w_{15}w_{46}w_{67}\phantom{w_{00}}\quad
 w_{26}w_{57}w_{67}\phantom{w_{00}}\quad
 w_{37}w_{57}w_{67}\phantom{w_{00}}                                \cr
}
$$
$n = 4$:
$$
\eqalign{
&w_{01}w_{03}w_{04}w_{06}\quad
 w_{01}w_{03}w_{06}w_{15}\quad
 w_{01}w_{04}w_{12}w_{24}\quad
 w_{01}w_{12}w_{15}w_{24}                                          \cr
&w_{01}w_{03}w_{17}w_{26}\quad
 w_{01}w_{12}w_{26}w_{35}\quad
 w_{01}w_{03}w_{17}w_{37}\quad
 w_{01}w_{12}w_{35}w_{37}                                          \cr
&w_{03}w_{04}w_{06}w_{45}\quad
 w_{03}w_{06}w_{15}w_{45}\quad
 w_{04}w_{12}w_{24}w_{45}\quad
 w_{12}w_{15}w_{24}w_{45}                                          \cr
&w_{03}w_{17}w_{26}w_{45}\quad
 w_{12}w_{26}w_{35}w_{45}\quad
 w_{03}w_{17}w_{37}w_{45}\quad
 w_{12}w_{35}w_{37}w_{45}                                          \cr
&w_{04}w_{06}w_{23}w_{47}\quad
 w_{06}w_{15}w_{23}w_{47}\quad
 w_{17}w_{23}w_{26}w_{47}\quad
 w_{17}w_{23}w_{37}w_{47}                                          \cr
&w_{04}w_{23}w_{24}w_{56}\quad
 w_{15}w_{23}w_{24}w_{56}\quad
 w_{23}w_{26}w_{35}w_{56}\quad
 w_{23}w_{35}w_{37}w_{56}                                          \cr
&w_{04}w_{06}w_{47}w_{67}\quad
 w_{06}w_{15}w_{47}w_{67}\quad
 w_{17}w_{26}w_{47}w_{67}\quad
 w_{17}w_{37}w_{47}w_{67}                                          \cr
&w_{04}w_{24}w_{56}w_{67}\quad
 w_{15}w_{24}w_{56}w_{67}\quad
 w_{26}w_{35}w_{56}w_{67}\quad
 w_{35}w_{37}w_{56}w_{67}.                                         \cr
}
$$
These are determined by $(A_2^3)_{ij}$ and $(A_2^4)_{ij}$, 
respectively, where $i,j \in \{(00,00),(01,01),$ $(10,10),(11,11)\}$, 
since this is the set of labels for the vertices in $G_2$ which lie 
in the subgraph isomorphic to $G_1$.

\vfill\eject
\global\setbox1=\hbox{[00]\enspace}
\parindent=\wd1

\noindent{\bf References}
\medskip

\parskip=0pt
%--------|---------|---------|---------|---------|---------|---------|
\item{[1]}
\feynman,
``Simulating physics with computers'',
\IJTP\ {\bf 21} (1982) 467--488.

\item{[2]}
L. Bombelli, J. Lee, D. A. Meyer and R. D. Sorkin,
``Spacetime as a causal set'', 
\PRL\ {\bf 59} (1987) 521--524;\hfb
\dajm,
``Spacetime Ising models'',
UCSD preprint (1995);\hfb
\dajm,
``Induced actions for causal sets'',
UCSD preprint (1995).

\item{[3]}
\feynman,
``Quantum mechanical computers'',
\FP\ {\bf 16} (1986) 507--531.

\item{[4]}
N. Margolus,
``Quantum computation'',
\ANYAS\ {\bf 480} (1986) 487--497.

\item{[5]}
C. S. Lent and P. D. Tougaw,
``Logical devices implemented using quantum cellular automata'',
\JAP\ {\bf 75} (1994) 1818--1825.

\item{[6]}
\gz,
``Quantum cellular automata'',
\CS\ {\bf 2}\break
(1988) 197--208.

\item{[7]}
S. Fussy, G. Gr\"ossing, H. Schwabl and A. Scrinzi,
``Nonlocal computation in quantum cellular automata'',
\PRA\ {\bf 48} (1993) 3470--3477.

\item{[8]}
\dajm,
``On the absence of homogeneous scalar quantum cellular automata'',
UCSD preprint (1995), quant-ph/9604011.

\item{[9]}
\dajm,
``From quantum cellular automata to quantum lattice gases'',
UCSD preprint (1995), quant-ph/9604003, to appear in \JSP

\item{[10]}
T. Toffoli and N. H. Margolus,
``Invertible cellular automata:  a review'',
\PD\ {\bf 45} (1990) 229--253.

\item{[11]}
K. Morita and M. Harao,
``Computation universality of one-dimensional reversible (injective)
  cellular automata'',
\TIEICEJE\ {\bf 72} (1989) 758--762.

\item{[12]}
J. Watrous,
``On one-dimensional quantum cellular automata'',
in 
{\sl Proceedings of the 36th Annual Symposium on Foundations of Computer 
Science}, Milwaukee, WI, 23--25 October 1995
(Los Alamitos, CA:  IEEE Computer Society Press 1995) 528--537.

\item{[13]}
E. Bernstein and U. Vazirani,
``Quantum complexity theory'',
in {\sl Proceedings of the 25th ACM Symposium on Theory of Computing},
San Diego, CA, 16--18 May 1993
(New York:  ACM Press 1993) 11--20.

\item{[14]}
\deutsch,
``Quantum theory, the Church--Turing principle and the universal
  quantum computer'',
\PRSLA\ {\bf 400} (1985) 97--117.

\item{[15]}
K. Morita,
``Computation-universality of one-dimensional one-way reversible
  cellular automata'',
\IPL\ {\bf 42} (1992) 325--329.

\item{[16]}
W. D. Hillis,
``New computer architectures and their relationship to physics or
  why computer science is no good'',
\IJTP\ {\bf 21} (1982) 255--262;\hfb
N. Margolus,
``Parallel quantum computation'',
in W. H. Zurek, ed.,
{\sl Complexity, Entropy, and the Physics of Information},
proceedings of the SFI Workshop, Santa Fe, NM, 
29 May--10 June 1989,
{\sl SFI Studies in the Sciences of Complexity} {\bf VIII}
(Redwood City, CA:  Addison-Wesley 1990) 273--287;\hfb
\brosl,
``Parallel billiards and monster systems'',
in N. Metropolis and G.-C. Rota, eds.,
{\sl A New Era in Computation}
(Cambridge:  MIT Press 1993) 53--65;\hfb
M. Biafore,
``Cellular automata for nanometer-scale computation'',
\PD\ {\bf 70}\break
(1994) 415--433;\hfb
R. Mainieri,
``Design constraints for nanometer scale quantum computers'',
preprint (1993) LA-UR 93-4333, cond-mat/9410109.

\item{[17]}
\teich\ and G. Mahler,
``Stochastic dynamics of individual quantum systems:  stationary
  rate equations'',
\PRA\ {\bf 45} (1992) 3300--3318;\hfb
A. Barenco, D. Deutsch, A. Ekert and R. Jozsa,
``Conditional quantum dynamics and logic gates'',
\PRL\ {\bf 74} (1995) 4083-4086;\hfb
J. I. Cirac and P. Zoller,
``Quantum computations with cold trapped ions'',
\PRL\ {\bf 74} (1995) 4091--4094.

\item{[18]}
C. D\"urr, H. L. Thanh and M. Santha,
``A decision procedure for well-formed linear quantum cellular 
  automata'',
in C. Puecha and R. Reischuk, eds.,
{\sl STACS 96:  Proceedings of the 13th Annual Symposium on 
     Theoretical Aspects of Computer Science},
Grenoble, France, 22--24 February 1996,
{\sl Lecture notes in computer science} {\bf 1046}
(New York:  Springer-Verlag 1996) 281--292.

\item{[19]}
C. D\"urr and M. Santha,
``A decision procedure for unitary linear quantum cellular
automata'',
preprint (1996) quant-ph/9604007.

\item{[20]}
P. A. M. Dirac,
{\sl The Principles of Quantum Mechanics}, fourth edition
(Oxford:  Oxford University Press 1958).

\item{[21]}
A. W. Burks,
``Von Neumann's self-reproducing automata'',
in A. W. Burks, ed.,
{\sl Essays on Cellular Automata}
(Urbana, IL:  University of Illinois Press 1970) 3--64.

\item{[22]}
S. Ulam,
``Random processes and transformations'',
in L. M. Graves, E. Hille, P. A. Smith and O. Zariski, eds.,
{\sl Proceedings of the International Congress of Mathematicians},
Cambridge, MA, 30 August--6 September 1950
(Providence, RI:  AMS 1952) {\bf II} 264--275;\hfb
J. von Neumann,
{\sl Theory of Self-Reproducing Automata},
edited and completed by A. W. Burks
(Urbana, IL:  University of Illinois Press 1966).

\item{[23]}
S. Wolfram,
``Computation theory of cellular automata'',
\CMP\ {\bf 96} (1984) 15--57.

\item{[24]}
H. Weyl,
{\sl The Theory of Groups and Quantum Mechanics},
translated from the second (revised) German edition by H. P. 
 Robertson
(New York:  Dover 1950).

\item{[25]}
P. R. Halmos,
{\sl A Hilbert Space Problem Book}, 
second edition, revised and enlarged
(New York:  Springer-Verlag 1982) Problems 52 and 127.

\item{[26]}
\rds,
``On the role of time in the sum-over-histories framework for 
  gravity'', 
presented at the conference on The History of Modern Gauge 
  Theories, Logan, Utah, July 1987,
published in \IJTP\ {\bf 33} (1994) 523--534;\hfb
\rds,
``Problems with causality in the sum-over-histories framework for
  quantum mechanics'',
in A. Ashtekar and J. Stachel, eds.,
{\sl Conceptual Problems of Quantum Gravity}, proceedings of the
  Osgood Hill Conference, North Andover, MA, 15--19 May 1988
(Boston:  Birkh\"auser 1991) 217--227;\hfb
J. B. Hartle,
``The quantum mechanics of closed systems'',
in B.-L. Hu, M. P. Ryan and C. V. Vishveshwara, eds.,
{\sl Directions in General Relativity:  Proceedings of the 1993
  international symposium, Maryland.  Volume 1:  papers in honor
  of Charles Misner\/}
(Cambridge: Cambridge University Press 1993) 104--124;\hfb
and references therein.

\item{[27]}
N. G. de Bruijn,
``A combinatorial problem'',
\PNAW\ {\bf 49} (1946) 758--764;\hfb
I. J. Good, 
``Normal recurring decimals'',
\JLMS\ {\bf 21} (1946) 167--169.

\item{[28]}
R. P. Stanley,
{\sl Enumerative Combinatorics}
(Monterey, CA:  Wadsworth \& Brooks/Cole 1986) Section 4.7.

\item{[29]}
J. E. Hopcroft and J. D. Ullman,
{\sl Introduction to Automata Theory, Languages, and Computation}
(Reading, MA:  Addison-Wesley 1979).

\item{[30]}
R. E. Blahut,
{\sl Theory and Practice of Error Control Codes\/}
(Reading, MA:  Addison-Wesley 1984) Section 14.3.

\item{[31]}
G. H. Mealy,
``A method for synthesizing sequential circuits'',
\BSTJ\ {\bf 34} (1955) 1045--1079.

\item{[32]}
E. F. Moore,
``Machine models of self-reproduction'',
\PSAM\ {\bf 14} (1962) 17--33;\hfb
J. Myhill,
``The converse of Moore's Garden-of-Eden Theorem'',
\PAMS\ {\bf 14} (1963) 685--686;\hfb
reprinted with revisions in A. W. Burks, ed.,
{\sl Essays on Cellular Automata}
(Urbana, IL:  University of Illinois Press 1970) 187--203; 204--205.

\item{[33]}
G. A. Hedlund,
``Endomorphisms and automorphisms of the shift dynamical system,''
\MST\ {\bf 3} (1969) 320--375.

\item{[34]}
D. Hillman,
``The structure of reversible one-dimensional cellular automata'',
\PD\ {\bf 52} (1991) 277--292.

\item{[35]}
H. Yamada and S. Amoroso,
``A completeness problem for pattern generation'',
\JCSS\ {\bf 4} (1970) 137--176.

\item{[36]}
S. Amoroso and Y. N. Patt,
``Decision procedures for surjectivity and injectivity of parallel
  maps for tesselation structures'',
\JCSS\ {\bf 6} (1972) 448--464.

\item{[37]}
W. Li, N. H. Packard and C. G. Langton,
``Transition phenomena in cellular automata rule space'',
\PD\ {\bf 45} (1990) 77--94;\hfb
and references therein.

\item{[38]}
N. Margolus,
``Physics-like models of computation'',
\PD\ {\bf 10} (1984) 81--95.

\item{[39]}
J. Kari,
``Reversibility and surjectivity problems of cellular automata'',
\JCSS\ {\bf 48} (1994) 149--182.

\item{[40]}
G. D. Doolen, U. Frisch, B. Hasslacher, S. Orszag and S. Wolfram, eds.,
{\sl Lattice Gas Methods for Partial Differential Equations},
a volume of lattice gas reprints and articles, incuding selected papers
from the Workshop on Large Nonlinear Systems, Los Alamos, NM,
August 1987,
{\sl SFI Studies in the Sciences of Complexity} {\bf IV}
(Redwood City, CA:  Addison-Wesley 1990);\hfb
H. Gutowitz,
{\sl Cellular Automata:  Theory and Experiment},
proceedings of a workshop sponsored by The Center for Nonlinear Studies,
Los Alamos National Laboratory, Los Alamos, NM,
9--12 September 1989,
reprinted from \PD\ {\bf 45} (1990)
(Amsterdam:  North-Holland 1990);\hfb
G. D. Doolen, ed.,
{\sl Lattice Gas Methods for PDE's:  Theory, Applications and Hardware},
proceedings of the NATO Advanced Research Workshop, Los Alamos, NM,
6--8 September 1989,
reprinted from \PD\ {\bf 47} (1991)
(Amsterdam:  North-Holland 1991).

\bye